\documentclass[lettersize,journal]{IEEEtran}

\usepackage{amsmath,amssymb,amsfonts, nccmath}
\usepackage{algorithmic}
\usepackage{array}
\usepackage[caption=false]{subfig}
\usepackage{textcomp}
\usepackage{stfloats}
\usepackage{mathtools}
\usepackage{enumitem}
\usepackage{url}
\usepackage{verbatim}
\usepackage{graphicx}
\usepackage{tabularx}
\usepackage[hidelinks]{hyperref}
\usepackage[most]{tcolorbox}
\usepackage[ruled,vlined]{algorithm2e}
\usepackage{soul}

\usepackage[capitalize,nameinlink]{cleveref}
\newcommand{\ballnumber}[1]{\tikz[baseline=(myanchor.base)] \node[circle,fill=.,inner sep=1pt] (myanchor) {\color{-.}\bfseries\footnotesize #1};}

\usepackage{todonotes}

\usepackage[acronyms,nonumberlist,nopostdot,nomain,nogroupskip,acronymlists={hidden}]{glossaries}

\usepackage[backend=bibtex,style=ieee,citestyle=numeric-comp,doi=false,isbn=false,maxnames=99,mincitenames=1,maxcitenames=2]{biblatex}
\usepackage{xcolor}
\def\BibTeX{{\rm B\kern-.05em{\sc i\kern-.025em b}\kern-.08em
    T\kern-.1667em\lower.7ex\hbox{E}\kern-.125emX}}

\usepackage{acro}
\newacronym{5g}{5G}{fifth-generation}
\newacronym{6g}{6G}{sixth-generation}
\newacronym{ti}{TI}{tactile internet}
\newacronym{aif}{AIF}{active inference}
\newacronym{vfe}{VFE}{variational free energy}
\newacronym{pomdp}{POMDP}{partially observable Markov decision process}
\newacronym{mdp}{MDP}{Markov decision process}
\newacronym{urllc}{URLLC}{ultra-reliable low-latency communication}
\newacronym{qoe}{QoE}{quality of experience}
\newacronym{qos}{QoS}{quality of service}
\newacronym{ai}{AI}{artificial intelligence}
\newacronym{bpsk}{BPSK}{binary phase-shift keying}
\newacronym{dnn}{DNN}{deep neural network}
\newacronym{dt}{DT}{digital twin}
\newacronym{gan}{GAN}{generative adversarial network}
\newacronym{genai}{GenAI}{generative artificial intelligence}
\newacronym{glfsr}{GLFSR}{Galois linear feedback shift register}
\newacronym{iq}{IQ}{in-phase and quadrature}
\newacronym{iqi}{IQI}{in-phase and quadrature-phase imbalance}
\newacronym{lstm}{LSTM}{long short-term memory}
\newacronym{nrmse}{NRMSE}{normalized root mean squared error}
\newacronym{ota}{OTA}{over-the-air}
\newacronym{sdr}{SDR}{software defined radio}
\newacronym{nextgen}{NextGWN}{next-generation wireless network}
\newacronym{rf}{RF}{radio frequency}
\newacronym{vae}{VAE}{variational autoencoders}
\newacronym{xai}{XAI}{explainable artificial intelligence}
\newacronym{drl}{DRL}{deep reinforcement learning}
\newacronym{rl}{RL}{reinforcement learning}
\newacronym{xrl}{XRL}{explainable reinforcement learning}
\newacronym{mpc}{MPC}{model predictive control}
\newacronym{efe}{EFE}{expected free energy}
\newacronym{xapp}{xApp}{eXtended application}
\newacronym{prb}{PRB}{physical resource block}
\newacronym{oai}{OAI}{OpenAirInterface}
\newacronym{embb}{eMBB}{enhanced mobile broadband}
\newacronym{mmtc}{mMTC}{massive machine-yype communications}
\newacronym{oran}{O-RAN}{open radio access network}
\newacronym{kpm}{KPM}{key performance measurement}
\newacronym{llm}{LLMs}{large language models}
\newacronym{ran}{RANs}{radio access networks}

\begin{document}

\title{BRAIN: Bayesian Reasoning via Active Inference for Agentic and Embodied Intelligence in Mobile Networks}

\author{%
    Osman Tugay Basaran,~\IEEEmembership{Student Member, IEEE},
    Martin Maier, and
    Falko Dressler,~\IEEEmembership{Fellow, IEEE}
\thanks{O.T. Basaran and Falko Dressler are with the School of Electrical Engineering and Computer
Science, TU Berlin, Berlin, 10587, Germany. E-mail: basaran@ccs-labs.org, dressler@ccs-labs.org}
\thanks{Martin Maier is with Optical Zeitgeist Laboratory, INRS, Montreal, QC H5A 1K6, Canada. E-mail: martin.maier@inrs.ca}
\thanks{This work has been funded by the Federal Ministry of Research, Technology and Space (BMFTR, Germany) as part of the technology transfer hub for the medicine and mobility of the future xG-RIC.}}



\maketitle

\begin{abstract}
Future \gls{6g} mobile networks will demand  \gls{ai} agents that are not only autonomous and efficient, but also capable of real-time adaptation in dynamic environments and transparent in their decision-making. However, prevailing agentic \gls{ai} approaches in networking, exhibit significant shortcomings in this regard. Conventional \gls{drl}-based agents lack explainability and often suffer from brittle adaptation, including catastrophic forgetting of past knowledge under non-stationary conditions. In this paper, we propose an alternative solution for these challenges: Bayesian reasoning via Active Inference ({\fontfamily{qcr}\selectfont BRAIN}) agent. {\fontfamily{qcr}\selectfont BRAIN} harnesses a deep generative model of the network environment and minimizes variational free energy to unify perception and action in a single closed-loop paradigm. We implement {\fontfamily{qcr}\selectfont BRAIN} as O-RAN \gls{xapp} on GPU-accelerated testbed and demonstrate its advantages over standard \gls{drl} baselines. In our experiments, {\fontfamily{qcr}\selectfont BRAIN} exhibits \textit{(i)} robust \emph{causal reasoning} for dynamic radio resource allocation, maintaining slice-specific \gls{qos} targets (throughput, latency, reliability) under varying traffic loads, \textit{(ii)} superior adaptability with up to 28.3\% higher robustness to sudden traffic shifts versus benchmarks (achieved without any retraining), and \textit{(iii)} real-time \emph{interpretability} of its decisions through human-interpretable belief state diagnostics. 
\end{abstract}

\begin{IEEEkeywords}
Active inference, Embodied-AI, mobile networks, trustworthiness, 6G.
\end{IEEEkeywords}

\glsresetall

\section{Introduction}\label{Intro}

\Gls{ai} has achieved remarkable advances in recent years, from mastering complex games and control tasks with \gls{rl} to producing human-like content with \gls{llm} and Generative \gls{ai}.
These achievements, however, remain largely \emph{disembodied}; models operate in simulated or data-driven domains without direct physical grounding.
\gls{llm}, for example, excel at pattern recognition and generation from static datasets but cannot interact with a changing environment.
Similarly, \gls{drl} agents typically train in carefully crafted simulations with fixed reward functions, and they often struggle when faced with real-world dynamics outside their training distribution.
In essence, today’s \gls{ai} systems lack the holistic, adaptive intelligence of an embodied agent that can continually perceive, act, and learn in the real world~\cite{liu2025embodied}.

\begin{figure}
    \centering
    \subfloat[Data distribution shifts during learning.]
    {%
        \includegraphics[width=\columnwidth]{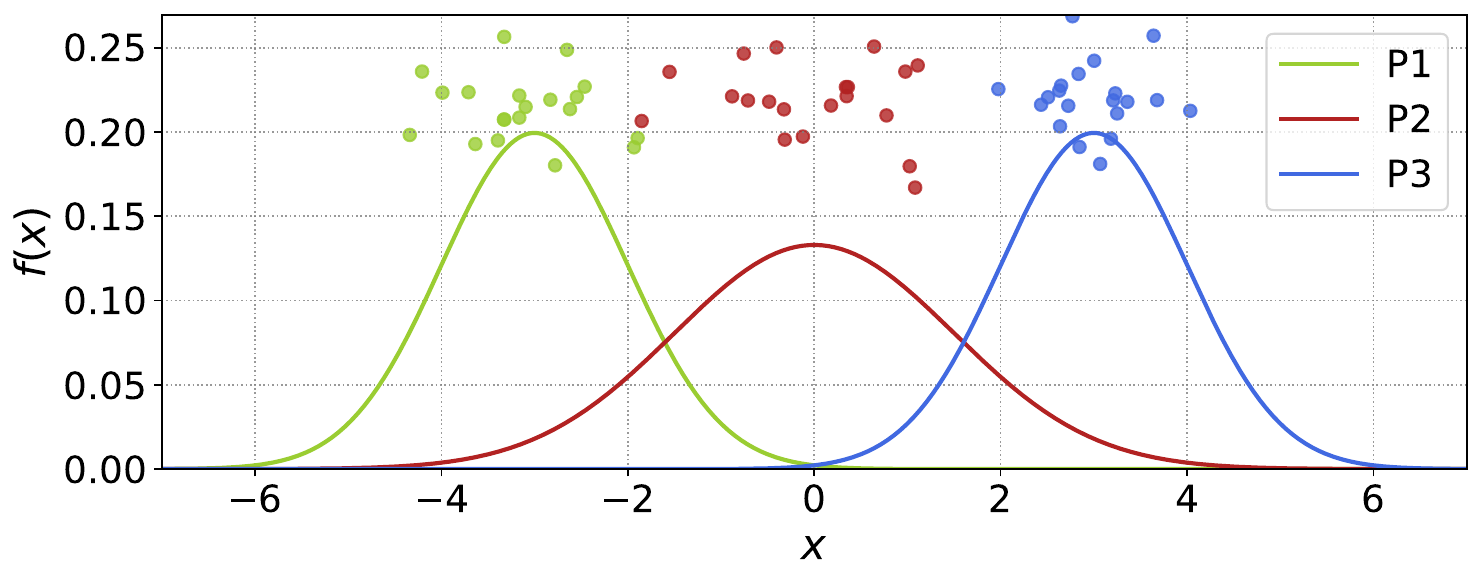}
    }
    \\
    \subfloat[Catastrophic forgetting.]{%
        \includegraphics[width=\columnwidth]{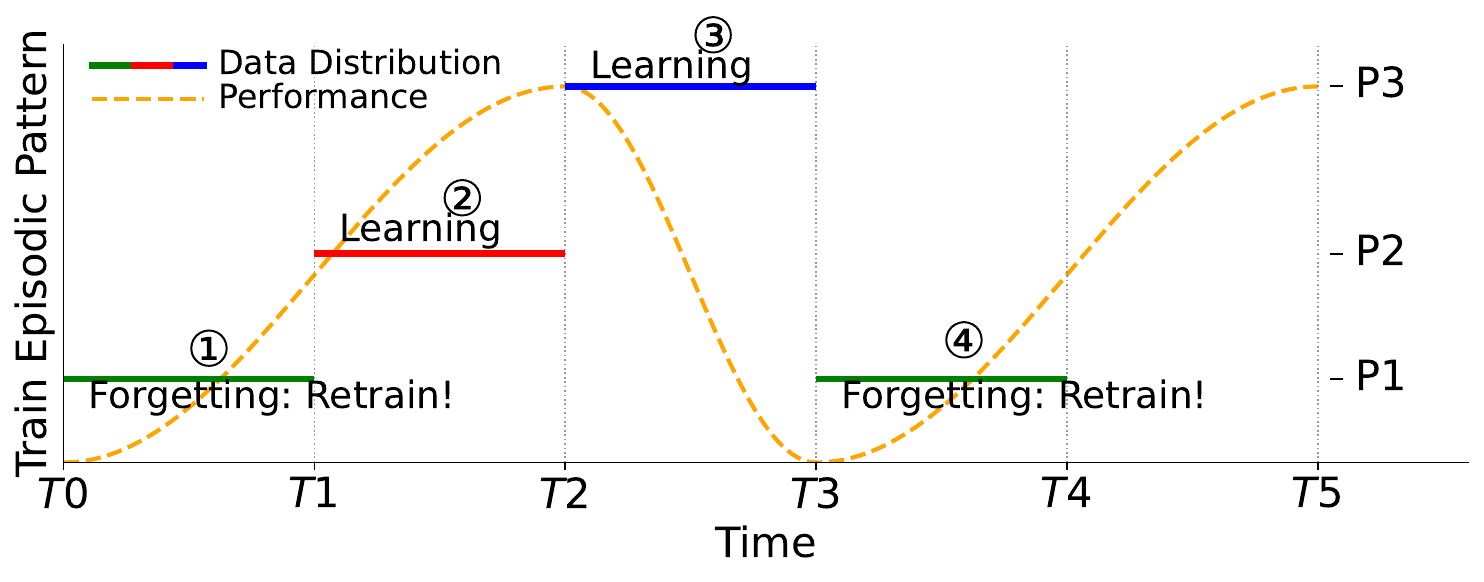}\label{fig_catast}
    }
    \caption{Preliminary experiment on catastrophic forgetting issue for \gls{drl} agent in network slicing. (a) Sequential distribution shifts: \gls{drl} agent trains initially on Slice 1 (eMBB, green), then shifts to Slice 2 (URLLC, red), and finally Slice 3 (mMTC, blue), each with distinct QoS requirements and traffic patterns. (b) The impact of distribution shifts on performance. Initially, at \textcircled{1} the agent learns Slice 1 effectively. When data distribution transitions at \textcircled{2} to Slice 2, the agent learns new policies while starting to forget previously acquired knowledge (Slice 1). At \textcircled{3}, another shift to Slice 3 occurs, further reducing performance on previously learned Slices 1 and 2. Eventually, at \textcircled{4}, the scenario cycles back, requiring costly retraining due to performance degradation caused by catastrophic forgetting. }
    \label{fig_preliminary}
    \vspace{-.8em}
\end{figure}

This gap becomes especially critical in the context of emerging \gls{6g} and beyond networks \cite{saad2025artificial}.
These future networks are expected to connect tens of billions of devices and support unprecedented services with stringent performance demands, necessitating \emph{AI-native} design principles that tightly integrate learning and control intelligence into the infrastructure \cite{saad2025artificial,polese2024empowering}.
The wireless environment is inherently complex and \emph{non-stationary}: channel conditions, user mobility, and traffic patterns fluctuate constantly \cite{cheng2022channel}.
Moreover, \gls{6g} must cater to a diverse array of \gls{qos}/\gls{qoe} requirements across use cases \cite{saad2020vision}.
Yet, most “AI-enabled” networking solutions to date simply apply off-the-shelf deep learning models (e.g., convolutional networks \cite{lecun1998convolutional} or deep autoencoders \cite{bank2021autoencoders}) to specific tasks, without fundamentally rethinking the network’s cognitive architecture \cite{thomas2024causal}.
While these models can learn mappings from historical data, they often fail to generalize when network conditions deviate from the training set. \Gls{rl} introduces a degree of agency by enabling \gls{ai} to learn through direct interaction with the environment \cite{sutton1998reinforcement}.
Indeed, DRL-based implementations have shown promise in wireless domains, tackling problems from dynamic spectrum allocation and power control to handoff optimization and end-to-end network slicing \cite{vincent2018introduction,liu2021onslicing,cheng2024oranslice}.
However, conventional \gls{drl} solutions suffer from two major shortcomings that limit their suitability as the “brains” of an autonomous \gls{6g} network.
First, \gls{drl} policies are typically realized by deep neural networks that act as opaque \emph{black boxes} \cite{vouros2022explainable}.
Second, standard \gls{drl} has very limited adaptability to changing conditions \cite{yang2025generalization}.
Once a DRL agent is trained for a given environment or traffic scenario, it tends to \emph{overfit} to those conditions.
Neural policies are prone to \textit{catastrophic forgetting}: when learning or fine-tuning on new data, they overwrite previously learned behaviors \cite{cheng2022channel}.
\Cref{fig_preliminary} illustrates this challenge in a network slicing scenario: a baseline \gls{drl} agent quickly “forgets” how to serve an \gls{embb} slice once it has adapted to an \gls{urllc} slice, and so on, necessitating expensive relearning for each recurrence of prior conditions.

These limitations point to the need for a fundamental reimagining of network \gls{ai}.
There is a growing consensus that next-generation networks should incorporate higher-level cognitive capabilities; integrating elements of memory, perception, and reasoning  rather than relying solely on low-level pattern recognition \cite{thomas2024causal,bariah2024ai}.
In essence, AI agents in such a system are no longer just offline models, but active participants in the physical network environment.
This agentic vision naturally leads to \emph{active inference as a promising next step for network intelligence.} Active inference has been described as an ideal framework for realizing such embodied AI, as it biomimics how natural intelligent systems learn and adapt via an action–perception loop grounded in the free energy principle \cite{maier2026from}.
Originally developed in cognitive neuroscience \cite{friston2016active}, active inference offers a unified theory of perception, learning, and action based on Bayesian reasoning.
In contrast to conventional \gls{rl}, an active inference agent does not rely on hand-crafted reward signals; instead, it maintains an internal \emph{generative model} of its environment and desired outcomes.
The agent constantly updates its beliefs about the hidden state of the world (perception) and selects actions to fulfill its goals by minimizing \emph{variational free energy}; a measure of prediction error or “surprise” between the agent’s expectations and its observations.
In essence, the agent tries to anticipate what should happen (given its model and goals) and then acts to make reality align with those expectations, thereby reducing surprise.

\begin{figure}
    \centering
    \subfloat[Traditional DRL Agent.]{\includegraphics[width=0.46\linewidth]{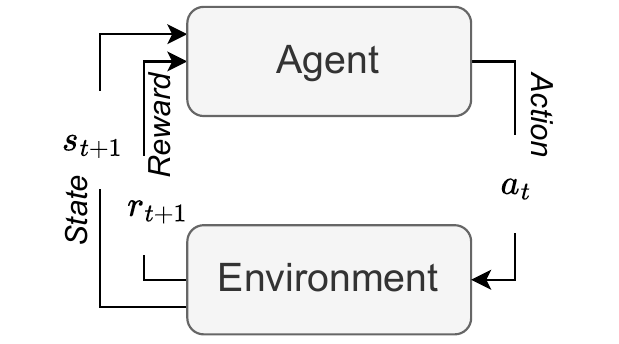}\label{fig_drlagent}}%
    \hfill
    \subfloat[Our {\fontfamily{qcr}\selectfont{BRAIN}} Agent.]{\includegraphics[width=0.54\linewidth]{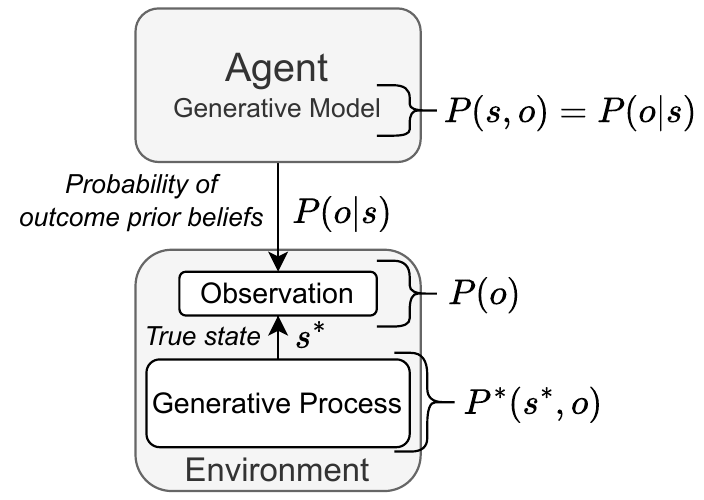}\label{fig_aif}}%
    \caption{Architecture of conventional DRL and proposed explainable deep active inference agents.}
    \label{fig_drvsac}
\end{figure}

In this paper, we introduce an explainable deep active inference agent for mobile network resource management on AI-RAN testbed. This work is a detailed and expanded version of a workshop paper currently under review.
Beyond the original core concept, we \textit{i)} broaden the experimental depth with additional advanced agent baselines and detailed ablations, \textit{ii)} add a controlled non-stationarity stress test across \emph{all} agents to quantify robustness and recovery, and \textit{iii)} include policy-entropy analysis to make exploration-exploitation dynamics comparable across \gls{drl} and active inference, alongside expanded sections and discussions. We call our framework {\fontfamily{qcr}\selectfont BRAIN} (\textbf{B}ayesian \textbf{R}easoning via \textbf{A}ctive \textbf{IN}ference), envisioning it as the “telecom brain” of an \gls{ai}-native RAN controller.
\Cref{fig_drvsac} contrasts the paradigm of a conventional DRL agent with that of our proposed {\fontfamily{qcr}\selectfont BRAIN} agent. The {\fontfamily{qcr}\selectfont BRAIN} architecture employs a deep generative active inference model to design the relationship between latent network states (e.g., congestion levels, channel conditions, user mobility) and observed performance metrics, while also encoding desired outcomes (e.g., slice-specific QoS targets) as prior beliefs.
At each control interval, the agent performs active inference by minimizing variational free energy: it infers the most probable current network state (perception step) and then computes the optimal resource allocation action (action step) that will drive the network’s predicted performance closer to the target (i.e., correcting the deviation between expected and desired outcomes).
This cycle of inference and action effectively allows {\fontfamily{qcr}\selectfont BRAIN} to carry out online learning and control simultaneously.
Unlike a DRL agent, which would require retraining whenever the environment changes, {\fontfamily{qcr}\selectfont BRAIN} continuously \emph{updates its beliefs in real time} as new observations arrive, endowing it with a form of lifelong learning that gracefully handles distribution shifts.
Moreover, because our agent’s internal computations revolve around probabilistic beliefs and free-energy contributions, we can tap into these intermediate results to understand and explain its behavior.

The core outcomes of our study are summarized as new contributions (“C") and new findings (“F") as follows:
\begin{itemize}
    \item[\textbf{C1.}] 
    We introduce {\fontfamily{qcr}\selectfont BRAIN}, the first deep active-inference agent for AI-RAN closed-loop RAN slicing in O-RAN.
    \item[\textbf{C2.}] 
    We designed {\fontfamily{qcr}\selectfont BRAIN} agent intrinsically explainable by exposing posterior beliefs over latent slice conditions and an \gls{efe} decomposition that justifies each action in terms of goal alignment (extrinsic) and uncertainty reduction (epistemic).
    
    \item[\textbf{F1.}] In dynamic slicing experiments, {\fontfamily{qcr}\selectfont BRAIN} demonstrates continual adaptation to non-stationary conditions. {\fontfamily{qcr}\selectfont BRAIN} more reliably maintains heterogeneous slice intents under dynamic loads. 
   \item[\textbf{F2.}]  Unlike black-box DRL, {\fontfamily{qcr}\selectfont BRAIN} exposes interpretable internal variables, enabling \emph{causal} and auditable explanations for resource-allocation decisions.
\end{itemize}

\section{Related Work}\label{RW}

This section situates {\fontfamily{qcr}\selectfont BRAIN} within three complementary research threads that underpin agentic intelligence in mobile networking. 
First, we review how \gls{rl}/\gls{drl} has been operationalized for network control and orchestration, particularly in \gls{oran} and slicing, as the dominant agentic paradigm in practice. 
Second, we summarize explainability efforts in wireless \gls{ai}, including \gls{xai} and emerging \gls{xrl} approaches, highlighting the extent to which interpretability is typically introduced \emph{post hoc} rather than being intrinsic to the decision process. 
Third, we discuss active inference as an embodied intelligence framework that unifies perception and action through probabilistic generative modeling and variational inference, and we identify the limited evidence to date on deploying \emph{deep} active inference with operator-facing explanations in communications systems. 
Together, these bodies of work clarify the methodological gap addressed by our approach: an \emph{intrinsically interpretable, continually adaptive agent} for real-time mobile network control.

\textbf{\Gls{rl} on Mobile Networks.} 
\Gls{rl} as well as \gls{drl} models have been increasingly adopted for dynamic resource management and control tasks in wireless networks.
Liu \textit{et al.} \cite{liu2021onslicing} propose OnSlicing, an online \gls{drl} framework for end-to-end network slicing across RAN, transport, core, and edge domains.
ORANSlice \cite{cheng2024oranslice}, an open source, modular platform for 5G network slicing tailored to the \gls{oran} ecosystem.
It integrates slice lifecycle management, resource orchestration, monitoring, and analytics within a flexible framework.
While RL policies can yield remarkable efficiency gains, their opaque nature often manifested as “black-box” neural networks hinders understanding and debugging, limiting practical deployment.
Therefore, recent research has begun to explore \gls{xrl} methodologies to improve transparency by explicitly elucidating policy decisions and learned behaviors. 

\textbf{\Gls{xai} on Mobile Networks.}
To overcome the transparency problem, researchers have turned to \gls{xai} techniques \cite{ali2023explainable, longo2024explainable} in the mobile networking domain \cite{guo2020explainable}.
In recent years, several works have explored using popular \gls{xai} methods (e.g., SHAP \cite{lundberg2017unified} and LIME \cite{ribeiro2016why}) to interpret complex models for wireless network tasks \cite{basaran2024xainomaly, 
 fiandrino2023explora, gizzini2024towards}.
While useful, such generic \gls{xai} approaches have proven insufficient for the needs of mobile networks. They provide only superficial insight and often struggle with the temporal and high-dimensional nature of network data. 
Recognizing these gaps, some studies have begun to pursue domain-specific \gls{xai} and inherently interpretable models for wireless communications.
Researchers introduced custom time-series explainers for network traffic models, which track how feature importance evolves over time and identify anomalous patterns leading to errors \cite{prez2025chronoprof, fiandrino2024aichronolens}.
In general, these efforts underline that explainability in wireless \gls{ai} may require expert-driven designs to meet the reliability and insight demands of network operations.
It is worth noting that \gls{xrl} is also gaining traction in other fields (like robotics and autonomous systems), aiming to extract human-comprehensible policies from \gls{rl} agents.
However, in the wireless networking literature, explainable \gls{rl} or \gls{drl} has seen very limited exploration so far.
One notable approach is SYMBXRL \cite{duttagupta2025symbxrl}, which introduces a symbolic explanation layer on top of black-box \gls{drl} models.
In this framework, a symbolic representation generator converts numerical state and action variables into discrete first-order logic predicates.
In contrast, our proposed framework adopts a fundamentally different paradigm by embedding explainability directly within the agent's generative and inferential processes.
Therefore, there remains a significant need for new methods that can interpret and justify the actions of different learning agents.

\textbf{Active Inference.}
In recent years, it has been applied in engineering domains, showing promise for state estimation, planning, and control under uncertainty \cite{dacosta2020Active}.
These early studies suggest that active inference can serve as a flexible, biologically inspired approach to sequential decision-making, distinct from  reinforcement learning \cite{tschantz2020reinforcement}. 
Note that RL formalisms for adaptive decision-making in unknown environments are subsumed by active inference.
Researchers have applied active inference to robot control tasks, where the agent's generative model enables it to handle ambiguous sensory inputs and still pursue goal-directed behavior \cite{oliver2022empirical, fujii2024realworld}. An intriguing aspect of active inference is its potential for built-in explainability, though this aspect has yet to be concretely validated \cite{maier2026from, albarracin2023designing}.

\section{Problem Formulation for Agentic AI Design}\label{Tech}

\subsection{Reinforcement Learning}

We model the closed-loop RAN slicing control problem as a sequential decision-making task under uncertainty, which can be formulated as a \gls{mdp}. The agent in our scenario is the Near-RT RIC control \gls{xapp}, and the environment consists of the gNB and its slices (\gls{embb}, \gls{urllc}, \gls{mmtc}) along with their traffic and radio conditions. At each discrete control interval $t$, the environment is in some state $s_t$ that captures the current slice performance, $s_t$ can be defined to include each slice’s downlink throughput, buffer occupancy, and transmitted transport block count (as obtained from \gls{oran} \gls{kpm} reports). The agent observes this state (or an observation derived from it) and then selects an action $a_t$ from its action space. In our use case, the action $a_t$ consists of setting the fraction of \glspl{prb} allocated to each slice and choosing a scheduling policy (e.g., PF, RR, WFQ) for each slice’s traffic. After the agent’s decision is applied, the environment transitions to a new state $s_{t+1}$ according to the network dynamics (influenced by the agent’s action and external factors such as user traffic and channel variability), and the agent receives a numerical reward $r_{t+1}$ reflecting the outcome.

The reward function $R(s_t, a_t, s_{t+1})$ is designed to quantify how well the slice-specific \gls{qos} objectives are met at time $t$. In particular, we assign positive reward for high \gls{embb} throughput, negative reward for large \gls{urllc} buffer occupancy (penalizing latency), and positive reward for successful \gls{mmtc} transmissions. Selected formulation is to define the immediate reward as a weighted sum of key performance metrics for the slices:
\begin{equation}\label{eq:reward}
\begin{multlined}
    r_{t+1} \;=\; \alpha \,\cdot\, \text{throughput}_{t+1}^{(\text{eMBB})} \;-\; \beta \,\cdot\, \text{buffer}_{t+1}^{(\text{URLLC})} \; \\ +\; \gamma \,\cdot\, \text{TBcount}_{t+1}^{(\text{mMTC})}\,,
\end{multlined}
\end{equation}
where $\text{throughput}_{t+1}^{(\text{eMBB})}$ is the \gls{embb} slice downlink data rate at time $t+1$, $\text{buffer}_{t+1}^{(\text{URLLC})}$ is the URLLC slice’s queue length (a proxy for its latency), $\text{TBcount}_{t+1}^{(\text{mMTC})}$ is the number of transport blocks successfully delivered for the mMTC slice, and $\alpha,\beta,\gamma$ are weighting coefficients reflecting the priority of each slice’s KPI. This reward design encourages the agent to maximize \gls{embb} throughput, minimize \gls{urllc} queuing delay, and ensure reliable \gls{mmtc} delivery, in line with the slices’ requirements.

The agent’s goal in \gls{rl} is to learn an optimal policy $\pi^*$ that maximizes the expected cumulative discounted reward. Starting from state $s_t$, the objective (also called return) can be written as: 
\begin{equation}\label{eq:rlagent}
G_t \;=\; \mathbb{E}\Bigg[\sum_{k=0}^{\infty} \gamma^k \, r_{t+k+1} \,\Big|\, s_t\Bigg]\,,
\end{equation}
with $\gamma \in [0,1)$ denoting a discount factor that prioritizes immediate rewards over future rewards. By maximizing this return, the \gls{rl} agent attempts to jointly satisfy the slices’ performance objectives over time (e.g., sustaining high throughput for \gls{embb}, low latency for \gls{urllc}, and robust throughput for \gls{mmtc}) even as network conditions evolve.

It is worth noting that the agent does not have direct access to all underlying network information (such as exact channel conditions or future traffic arrivals); it relies on the observed slice metrics as its state representation. In other words, the environment is partially observable from the agent’s perspective. We can view the problem as a \gls{pomdp}, where the \gls{oran} \gls{kpm} reports constitute the observation $o_t$ that provides a noisy, partial view of the true state $s_t$. In our design, we assume these reported metrics are a sufficient statistic of the network’s condition, and the agent (potentially using function approximation or recurrent networks) can internally maintain any additional state context needed.

\subsection{Active Inference}\label{ActiveInference}

Active inference is a novel decision-making paradigm stemming from cognitive neuroscience that offers a unified approach to action and perception under uncertainty \cite{parr2022active}. Instead of learning a policy purely from external reward feedback, an active inference agent leverages an internal generative model of its environment and acts so as to minimize the “surprise” (prediction error) of its observations. In our scenario, this means the agent (our {\fontfamily{qcr}\selectfont BRAIN} \gls{xapp}) is designed with prior expectations about the RAN slicing system; for example, that the URLLC slice’s buffer should remain low (to indicate low latency) and the eMBB slice’s throughput should be high. The agent then continuously adjusts its actions to align the observed slice performance with these internal expectations, thereby reducing unexpected deviations from desired behavior.

Mathematically, active inference casts the problem of closed-loop control as a variational inference process. The agent possesses a probabilistic \emph{generative model} of the RAN environment and treats the true network state as a latent (hidden) variable to be inferred. We can formalize this generative model over a time horizon $T$ by the joint distribution:
\begin{equation}\label{eq:actv}
P(o_{1:T},\,s_{0:T}) \;=\; P(s_0)\,\prod_{t=1}^{T} P(s_t \mid s_{t-1}, a_{t-1}) \; P(o_t \mid s_t)\,,
\end{equation}
where $s_t$ represents the hidden state of the network at time $t$ and $o_t$ is the observation (slice performance metrics) at time $t$. Here $P(s_0)$ is the prior distribution over the initial state, $P(s_t \mid s_{t-1}, a_{t-1})$ is the state transition model encoding the dynamics of the slices (how the true state evolves given the previous state and control action), and $P(o_t \mid s_t)$ is the observation likelihood model (mapping the hidden state to the probability of seeing a particular set of \gls{kpm} observations). In our context, $s_t$ may encompass the underlying radio and traffic conditions for each slice (e.g., actual backlog, channel quality, etc.), while $o_t$ includes the reported throughput, buffer occupancy, and TB count for each slice. The agent never directly observes $s_t$; instead it receives $o_t$ and must infer $s_t$ from these measurements.

To enable principled inference, the agent maintains an approximate posterior belief $q(s_t)$ over the hidden state and refines this belief with each new observation. The quality of the agent’s belief relative to the true state is quantified by the \emph{variational free energy} (VFE) \cite{friston2010free-energy}:
\begin{equation}\label{eq:genmodel}
\mathcal{F}(q, o_t) \;=\; \mathbb{E}_{q(s_t)}\!\Big[\,\ln q(s_t) \;-\; \ln P(o_t, s_t)\,\Big]\,,
\end{equation}
which measures how “surprising” the observation $o_t$ is under the agent’s current belief (it can be viewed as the negative evidence lower bound for the model). By minimizing $\mathcal{F}$ with respect to $q(s_t)$, the agent updates its posterior to better explain the observed data; in effect, performing a Bayesian update to reduce prediction error. This continual belief update (often implemented via gradient descent or closed-form Bayesian filtering) corresponds to the perception phase of active inference: the \gls{xapp} assimilates the latest \gls{kpm} measurements and adjusts its internal state estimate so that its predictions about slice performance become more accurate.

What truly distinguishes active inference from passive inference schemes is that the agent also \emph{acts} to minimize expected future surprise. In other words, beyond updating its beliefs, the agent plans and takes actions that it expects will lead to observations aligning with its preferences (and reducing uncertainty). This is formalized through the \gls{efe} of a policy $\pi$. A policy $\pi = \{a_t, a_{t+1}, \dots, a_{t+H}\}$ is a sequence of future actions of length $H$ (the planning horizon). The \gls{efe} for such a policy can be expressed as:
\begin{equation}\label{eq:gpi}
G(\pi) \;=\; \mathbb{E}_{q(o_{>t},\,s_{>t} \mid \pi)} \Big[\,\ln q(s_{>t} \mid o_{>t}) \;-\; \ln P(o_{>t}, s_{>t})\,\Big]\,,
\end{equation}
where $o_{>t}$ and $s_{>t}$ denote the future observations and states from time $t+1$ up to $T$ (under the hypothesis that policy $\pi$ is executed). Intuitively, $G(\pi)$ predicts how much “surprise” would be encountered if the agent were to follow policy $\pi$. Policies that are expected to produce observations close to the agent’s preferred outcomes (and that reduce uncertainty about the state) will have lower $G(\pi)$. Thus, the action selection rule in active inference is to choose the policy (or action, in a look-ahead setting) that minimizes $G(\pi)$. In our RAN slicing use case, this means the agent will favor actions that it believes will drive the slice metrics towards their ideal ranges;  keeping the \gls{urllc} buffer near empty and the \gls{embb} throughput high; while also gathering information to improve its state estimates when needed. In effect, the agent balances exploitation and exploration automatically: actions are chosen both to fulfill slice \gls{qos} goals (according to the agent’s internal goal preferences) and to resolve significant uncertainties (if the agent is unsure about some aspect of the network state).

\begin{tcolorbox}[colback=yellow!10!white,colframe=gray!75!black,title=Generative Model]
\ballnumber{1} \textbf{States ($S$):} Hidden state variables $s \in S$ representing the true underlying condition of the RAN slices. This can include factors like actual traffic load and channel quality for each slice, which are not directly observed but influence performance. The hidden state at time $t$ mediates the generation of observations $o_t$.  \ballnumber{2} \textbf{Observations ($O$):} Measurable outputs $o \in O$ that the agent can perceive from the environment. In our case, these are the slice-level \gls{kpm}s (throughput, buffer occupancy, TB count per slice) reported at each control interval. In a fully observable setting one might have $o_t = s_t$, but here $o_t$ provides a partial, noisy view of $s_t$. \ballnumber{3} \textbf{Actions ($U$):} Control inputs $u \in U$ that the  {\fontfamily{qcr}\selectfont BRAIN} agent can take to influence the state. For the slicing scenario, an action defines how resources are allocated to slices (PRB fractions) and which scheduling policy is applied, thereby affecting how the state $s$ evolves (e.g., altering how quickly a slice’s queue is drained or how throughput is shared). \ballnumber{4} \textbf{Observation Model ($A$):} The probabilistic mapping from states to observations. $A$ specifies the likelihood of each possible observation given a state, $A_{o,s} = P(o \mid s)$. In our model, $A$ captures how a particular network state (with certain true throughput demand, backlog, channel conditions, etc.) would probabilistically produce the observed \gls{kpm} metrics. This includes any measurement noise or uncertainty in observation. \ballnumber{5} \textbf{Transition Model ($B$):} The state transition dynamics under actions. $B$ is defined such that $B_{s',s}^{(u)} = P(s_{t} = s' \mid s_{t-1}=s,\,u_{t-1}=u)$, describing the probability of the state moving from $s$ to $s'$ given that action $u$ was taken. In the RAN slicing context, $B$ encodes the effect of control actions on the network state. 
\end{tcolorbox}

In summary, our active inference-based controller continuously updates its internal model of the RAN slices and selects resource control actions that minimize the expected free energy. This leads to a closed-loop behavior wherein the agent seeks to make its observations unsurprising by ensuring slice performance meets the target objectives. Notably, this framework naturally handles partial observability (treating the true network conditions as latent variables to be inferred) and accommodates multiple slice objectives through built-in preference encoding (each slice’s QoS target is reflected in the agent’s model as a preferred outcome). The result is a principled control strategy that, unlike standard RL, does not require an externally defined reward function for each scenario but rather emerges from the agent’s intrinsic drive to minimize prediction error and uphold its modeled service goals.

\section{Explainable Deep Active Inference Design}\label{XDAI}

\subsection{Generative Model Design}\label{GEN}

At the core of our framework is a generative model that describes the network slicing environment without hierarchical abstractions.
We define a set of hidden state variables $s_t$ that characterize the real-time status of the network and its slices at time $t$.
$s_t$ include latent features such as per-slice traffic load, channel quality indicators, or queue lengths for each network slice.
The agent receives observations $o_t$ (e.g., measured throughput, latency, or slice performance metrics) that are probabilistically generated from the hidden state.
The agent can also perform actions $a_t$ (or controls) corresponding to slice resource allocation decisions. Without loss of generality, we refer to scenarios such as distributing a limited pool of \gls{prb} among slices or selecting scheduling policies for each slice, as typically encountered in O-RAN systems. In \cref{alg:brain-loop},
{\fontfamily{qcr}\selectfont{BRAIN}}'s generative model is expressed as a joint distribution over states, observations, and actions across time, $P(s_{0:T}, o_{0:T}, a_{0:T})$, which factorizes into dynamical and observational components:
%
\begin{equation}
\begin{multlined}
    P(s_{0:T}, o_{0:T}, a_{0:T}) = \\ P(s_0)\prod_{t=0}^{T-1} P(o_t \mid s_t),P(s_{t+1} \mid s_t, a_t),P(a_t). 
\end{multlined}
\end{equation}

\begin{algorithm}[t]
\DontPrintSemicolon
\caption{Proposed {\fontfamily{qcr}\selectfont{BRAIN}} Agent}
\label{alg:brain-loop}
\KwIn{Generative model $P(s_{t+1}, o_{t+1} \mid s_t, a_t)$; Preference distribution $P_{\text{pref}}(o)$; Action set $\mathcal{A}$; Time horizon $T$; Prior belief $Q(s_0)$.}
\KwOut{Selected action sequence $\{a_0, a_1, \ldots, a_{T-1}\}$; Logged beliefs and free-energy terms for explainability.}
Initialize $Q(s_0)$ (prior belief over states)\;
\For{$t \leftarrow 0$ \KwTo $T-1$}{
    $o_t \leftarrow$ observe new data from environment\;\tcp*[r]{Receive observation $o_t$ (e.g., current network slice metrics)}
    $Q(s_t) \leftarrow \text{BayesianUpdate}(Q(s_{t-1}), o_t)$\;\tcp*[r]{Update posterior state belief given $o_t$}
    \tcp{Belief update yields $Q(s_t)$ }
    \ForEach{$a_t \in \mathcal{A}$}{
        Compute $Q(o_{t+1} \mid a_t)$ using generative model\; \tcp*[r]{Predict observation distribution if action $a_t$ is taken}
        $\text{KL}_{\text{pref}} \gets D_{\mathrm{KL}}\!\Big( Q(o_{t+1}\mid a_t)\,\Big\|\, P_{\text{pref}}(o_{t+1}) \Big)$\; \tcp*[r]{Divergence from preferred outcomes (extrinsic term)}
        $I_{\text{gain}} \gets I(s_{t+1};\, o_{t+1} \mid a_t)$\;\tcp*[r]{Expected information gain (epistemic term)}
        $G(a_t) \gets \text{KL}_{\text{pref}} - I_{\text{gain}}$\;\tcp*[r]{Expected free energy $G(a_t)$}
    }
    $a_t^* \gets \arg\min_{a_t \in \mathcal{A}} G(a_t)$\;\tcp*[r]{Select action that minimizes expected free energy}
    Execute $a_t^*$ on environment\; \tcp*[r]{Apply chosen action (Adjust network slice resources)}
    Log $Q(s_t)$, $\{G(a) : a \in \mathcal{A}\}$, and $a_t^*$ for analysis\; \tcp*[r]{Record beliefs and free-energy terms for explainability}
}
\end{algorithm}

\subsection{Variational Inference and Policy Selection via Free Energy Minimization}

\textit{Perception as Variational Inference:} Upon receiving new observation $o_t$ (e.g., current slice throughput levels), the agent must infer the latent state $s_t$ (e.g., actual demand or user load causing those throughput levels).
{\fontfamily{qcr}\selectfont{BRAIN}} performs this by minimizing the variational free energy $F_t$, which serves as a proxy for Bayesian inference. We define the free energy at time $t$ as:
%
\begin{equation}\label{free}
\begin{multlined}
F_t(Q(s_t)) = \mathbb{E}{Q(s_t)}[-\ln P(o_t \mid s_t)-\\ 
 \ln P(s_t \mid s_{t-1},a_{t-1})]  + \mathbb{H}[Q(s_t)],
\end{multlined}
\end{equation}
%
where $\mathbb{H}[Q]$ is the entropy of the approximate posterior $Q$. Intuitively, minimizing $F_t$ encourages the posterior $Q(s_t)$ to explain the observation well (high likelihood $P(o_t\mid s_t)$) while staying close to the prior prediction $P(s_t \mid s_{t-1},a_{t-1})$ (which is the generative model's forecast from the previous time, ensuring temporal consistency).
In practice, we assume a tractable form for $Q(s_t)$ (i.e., Gaussian distribution or delta at a point estimate) and perform a gradient-based update (or closed-form update if linear-Gaussian) to find $Q^*(s_t) \approx P(s_t \mid o_{\le t}, a_{<t})$.
This posterior belief $Q(s_t)$ encapsulates the agent's current understanding of network conditions after seeing the data.
Notably, because the generative model is explicit, this belief is fully transparent: each variable in $s_t$ (say, the estimated load on slice 1) has a quantitative posterior that can be reported or visualized. The ability to inspect $Q(s_t)$ at any time means that the agent's situational awareness is interpretable and auditable.

\subsection{Action Selection as Expected Free Energy Minimization}

After updating its beliefs, the agent must decide on the next action $a_t$ (e.g., how to reallocate resources among slices) before the next observation arrives.
Instead of using a learned Q-value or policy network as in RL, {\fontfamily{qcr}\selectfont{BRAIN}} evaluates prospective actions by their expected free energy $G(\pi)$, where $\pi$ denotes a candidate policy (a sequence of future actions, or simply the single action $a_t$ in a one-step horizon case).
The agent chooses the policy that minimizes $G(\pi)$, reflecting the principle of active inference: actions are chosen to minimize expected surprise and fulfill preferences.
We formulate $G(\pi)$ (for simplicity, with a one-step horizon $\pi=a_t$) as:
\begin{equation}\label{expected}
\begin{multlined}
G(a_t) = \mathbb{E}{Q(s_{t+1}\mid s_t, a_t)}\Big[ -\ln P(o_{t+1} \mid C)+ \\ 
\mathrm{D_{KL}}\big(Q(s_{t+1}\mid o_{t+1}) | Q(s_{t+1})\big) \Big],
\end{multlined}
\end{equation}
where the expectation is taken with respect to the predicted next-state distribution under action $a_t$, and inside we consider two quantities for the next time step $t+1$:
\begin{enumerate}                [label=\large\protect\textcircled{\small\arabic*}] 

\item Expected “Surprise” w.r.t Preferences:
$-\ln P(o_{t+1} \mid C)$ is the surprisal (negative log-probability) of a future observation $o_{t+1}$ under the preference model.
If an action is likely to produce outcomes aligned with $C$ (e.g., all slices meet their targets), this term will be low; if the action would result in poor performance for some slice, yielding an observation far from the preferred range, this term will be high.
Thus, the expected value $\mathbb{E}[-\ln P(o_{t+1}\mid C)]$ serves as a risk or phenotypic cost for that action.

\item Expected Information Gain (Epistemic Value):
The second term $\mathrm{D_{KL}}[Q(s_{t+1}\mid o_{t+1}) | Q(s_{t+1})]$ is the $\mathrm{KL}$ divergence between the posterior and prior predicted state at $t+1$, which quantifies how much we expect to learn about the hidden state by observing $o_{t+1}$.
Equivalently, it can be expressed as the mutual information $I[s_{t+1}; o_{t+1}]$ between state and observation under that action.
An action that is expected to resolve uncertainty (e.g., probing an uncertain slice's condition by allocating it resources to see if throughput improves) will have a high information gain, reducing $G$.
This corresponds to the perceptual ambiguity term in active inference, encouraging exploratory actions that reduce uncertainty.

\end{enumerate}

\subsection{Introspective Explainability of Decisions}

At each time $t$, the {\fontfamily{qcr}\selectfont{BRAIN}} agent maintains a posterior belief distribution over the latent slice states $s_t$ (e.g., each slice's current demand level or reliability). We denote this belief as:
\begin{equation}\label{exp}
Q(s_t) = P(s_t \mid o_{1:t}, a_{1:t-1}),
\end{equation}
with the probability of each hidden state given all past observations $o_{1:t}$ and actions $a_{1:t-1}$.
In practice, $Q(s_t)$ is computed via the agent's variational Bayes update after receiving observation $o_t$.
For example, if slice demand can be high or low, $Q(s_t)$ might be a probability ${P(s_t=\text{high})}$ or ${P(s_t=\text{low})}$ that is updated as new traffic measurements come in.
These posterior beliefs are introspective variables because they represent the agent's internal knowledge about the network (and can be exposed for explainability).
The distribution $Q(s_t)$ is normalized and reflects the agent's confidence in different slice conditions at time $t$. 

The agent encodes prior preferences about desirable outcomes for the network slices. Let $P_{\mathrm{pref}}(o)$ denote the preferred distribution over observations (or performance metrics). $P_{\mathrm{pref}}(o)$ could assign high probability to outcomes where each slice's throughput or latency meets its target.
At each decision step, the agent predicts the distribution of next observations $Q(o_{t+1}\mid a_t)$ for a candidate action $a_t$ (by marginalizing over its state beliefs).
We define a preference alignment cost as the Kullback–Leibler (KL) divergence between the predicted outcome distribution and the preferred distribution. Formally, for action $a_t$:
\begin{equation} \label{exp2}
\begin{multlined} 
D_{\mathrm{KL}}\Big( Q(o_{t+1}\mid a_t)\Big|P_{\mathrm{pref}}(o_{t+1}) \Big) = \\ \sum_{t=0}^{\infty} Q(o_{t+1}\mid a_t)\ln \frac{Q(o_{t+1}\mid a_t)}{P_{\mathrm{pref}}(o_{t+1})}.
\end{multlined}
\end{equation}

\begin{figure}
  \centering
  \includegraphics[width=.9\columnwidth]{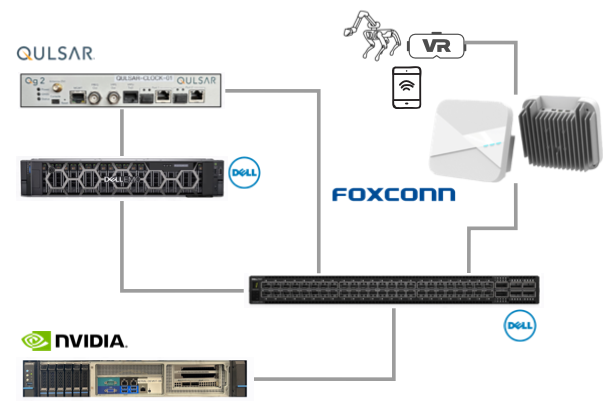}
  \caption{Overview of GPU-Accelerated Testbed.}
  \label{fig:testbed}
  \vspace{-.8em}
\end{figure}

This term measures how well the agent's expected observations align with its preferences (a form of phenotypic risk).
A lower KL value means the predicted network performance under action $a_t$ is closer to the ideal (preferences), indicating good extrinsic alignment with the slice QoS goals.
A high value indicates a misalignment; particularly the agent expects outcomes (like a slice throughput shortfall) that deviate from what is desired.
Minimizing this term drives the agent to choose actions that fulfill slice requirements (extrinsic reward-seeking behavior).
In parallel, the agent quantifies the epistemic value of actions, i.e., the expected information gain about the latent slice states.
This captures the action's exploratory benefit: how much taking action $a_t$ will reduce uncertainty in $s_{t+1}$.
Mathematically, we can define the epistemic value as the mutual information between future states and observations, conditioned on $a_t$.
One convenient form is the expected reduction in entropy of the state-belief after observing $o_{t+1}$:
\begin{equation}
\begin{multlined}
I(s_{t+1},o_{t+1}\mid a_t) = H\big[Q(s_{t+1}\mid a_t)\big]- \\ \mathbb{E}{Q(o_{t+1}\mid a_t)}\Big[H\big[Q(s_{t+1}\mid a_t,o_{t+1})\big]\Big], 
\end{multlined}    
\end{equation}
where $H[Q(s)] = -\sum_s Q(s)\ln Q(s)$ is the entropy.
Intuitively, $H[Q(s_{t+1}\mid a_t)]$ is the agent's prior uncertainty about the next state before taking action $a_t$, and the second term is the expected posterior uncertainty after seeing the new observation.
Their difference is the expected information gain.
If action $a_t$ is purely informational (i.e., a probing action that reveals a slice's condition), this quantity will be high, meaning the agent anticipates learning a lot (large epistemic value).
If an action is not informative, i.e., it does not affect observations, the epistemic value is low.
In many active inference formulations, this corresponds to reducing \textit{perceptual ambiguity}.

\section{Experiment Design}

\subsection{GPU-Accelerated AI-RAN Testbed}

We deploy a private 5G testbed (see Figure \ref{fig:testbed}) featuring a GPU-accelerated \gls{oran} architecture built on the NVIDIA Aerial Research Cloud (ARC) platform \cite{kelkar2021aerial, kelkar2021nvidia} and Aerial SDK \cite{nvidiaaerial}.
In our setup, the gNB's protocol stack is split into an O-DU Low (Layer-1 PHY) running on an NVIDIA GPU and an O-DU High/CU (higher layer protocols) running on $x86$ CPUs with \gls{oai} \cite{kaltenberger2020openairinterface}.
The two halves communicate via the Small Cell Forum's FAPI interface, enabling inline acceleration of PHY-layer DSP tasks on the GPU while maintaining a standard OAI software stack for MAC/RLC/PDCP/RRC.
Foxconn O-RU \cite{foxconn} operating in the n78 TDD band (mid-band FR1) provides the radio front-end, connected over a standard O-RAN 7.2 fronthaul interface.
This O-RU supports a 100 MHz channel bandwidth (273 \gls{prb}s at 30 kHz subcarrier spacing) in TDD mode, with a TDD pattern configured according to 3GPP Release 15 specifications (e.g., DDDSU slots). The testbed is equipped with both commercial and softwarized UEs to generate multi-slice traffic.
In particular, we use a COTS 5G UE (Sierra Wireless EM9191 modem module) and an \gls{oai}-based soft UE (nrUE) as two end devices.

The Sierra Wireless EM9191 provides a real 5G NR user equipment that connects over-the-air to the gNB.
The \gls{oai} nrUE is a software UE stack (also running on a server with an SDR front-end) that emulates a second 5G UE, allowing fine-grained control of its traffic and slicing configuration.
Both UEs support establishing multiple PDU sessions concurrently, which we map to different network slices (as described next).

\begin{figure*}
    \centering
    \subfloat[Mean cumulative reward.]{%
        \includegraphics[width=0.30\linewidth]{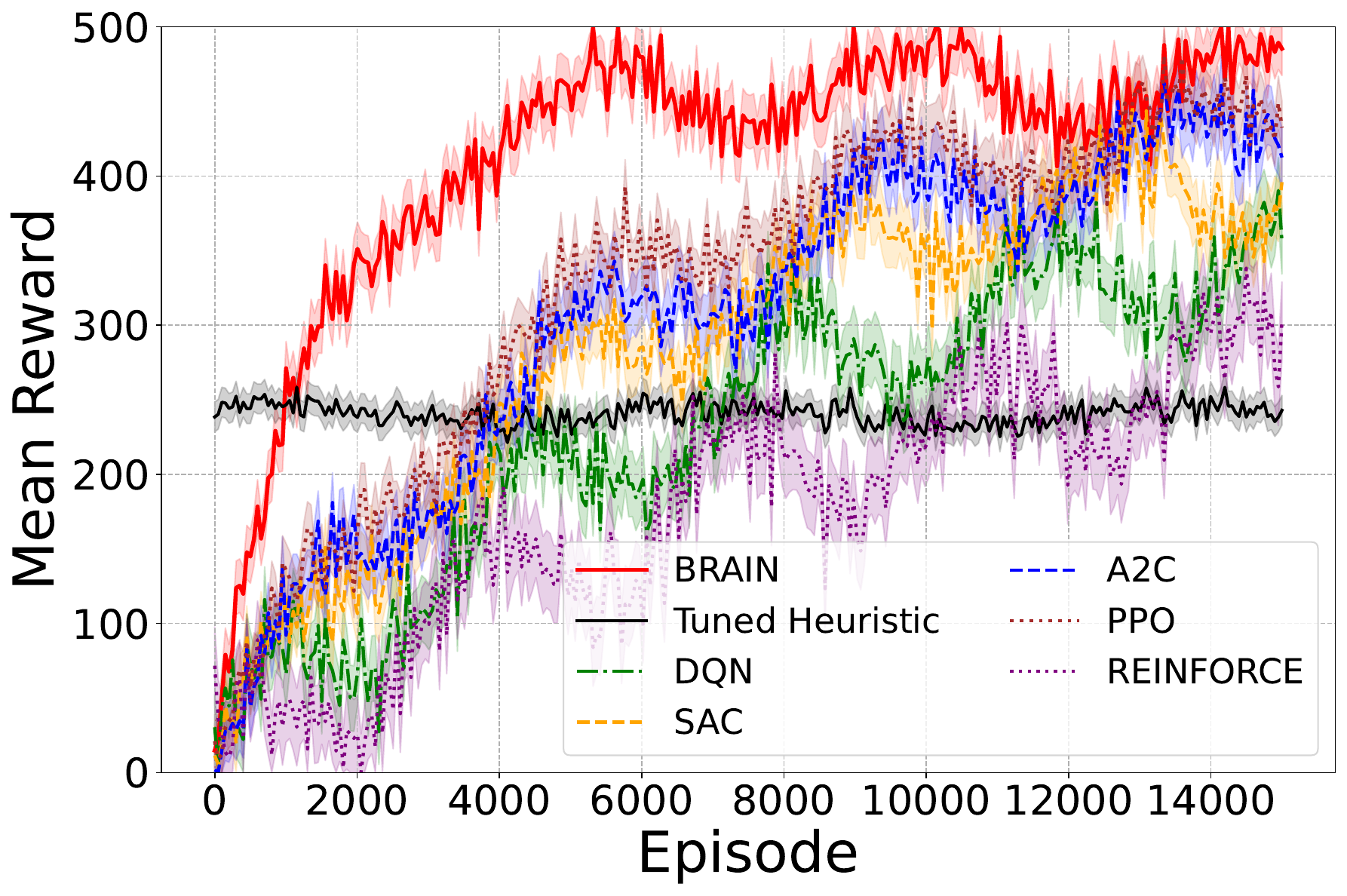}%
        \label{fig_reward}%
    }%
    \hfill
    \subfloat[Average loss (stability).]{%
        \includegraphics[width=0.30\linewidth]{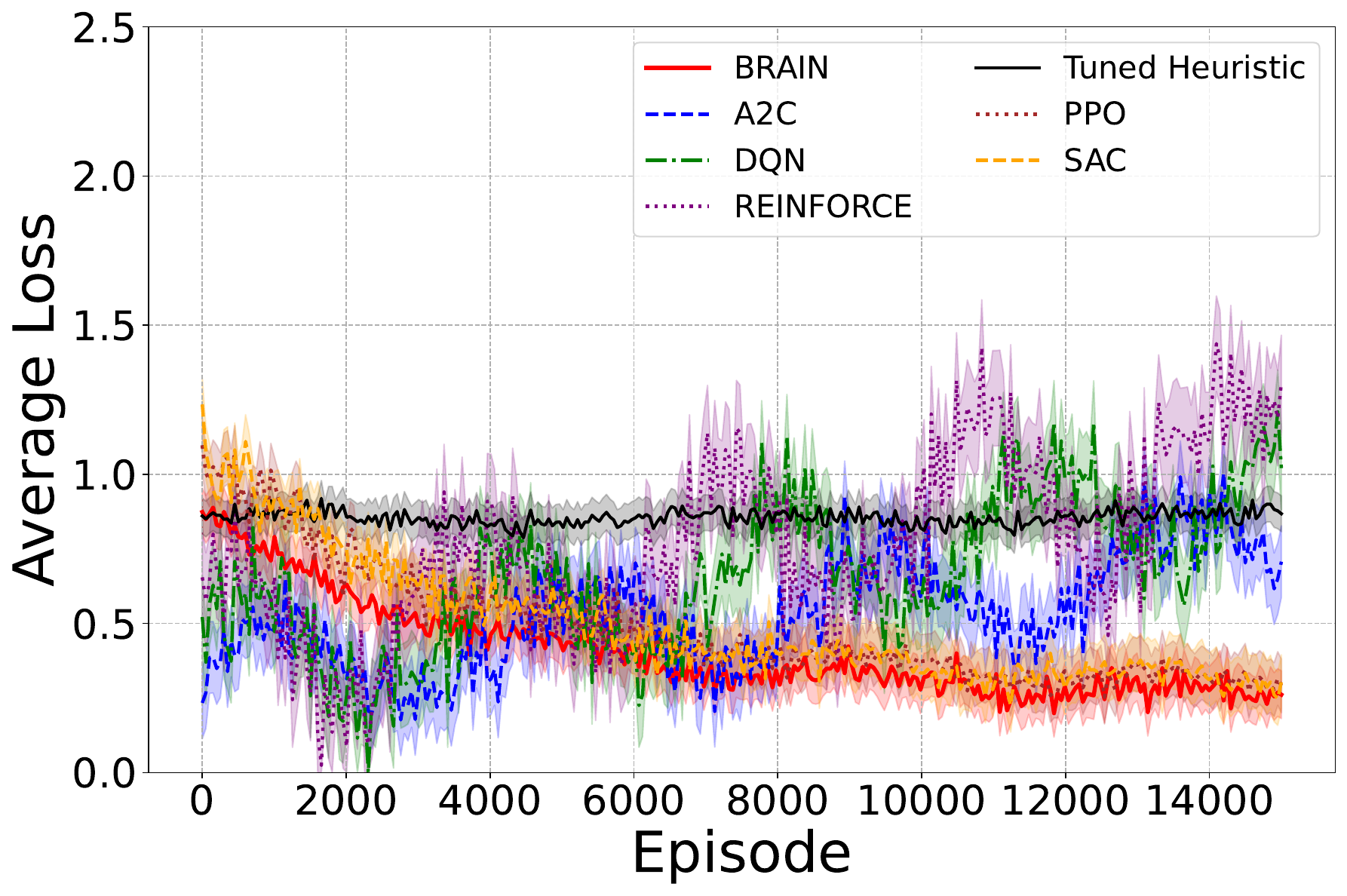}%
        \label{fig_loss}%
    }%
    \hfill
    \subfloat[Policy entropy.]{%
        \includegraphics[width=0.30\linewidth]{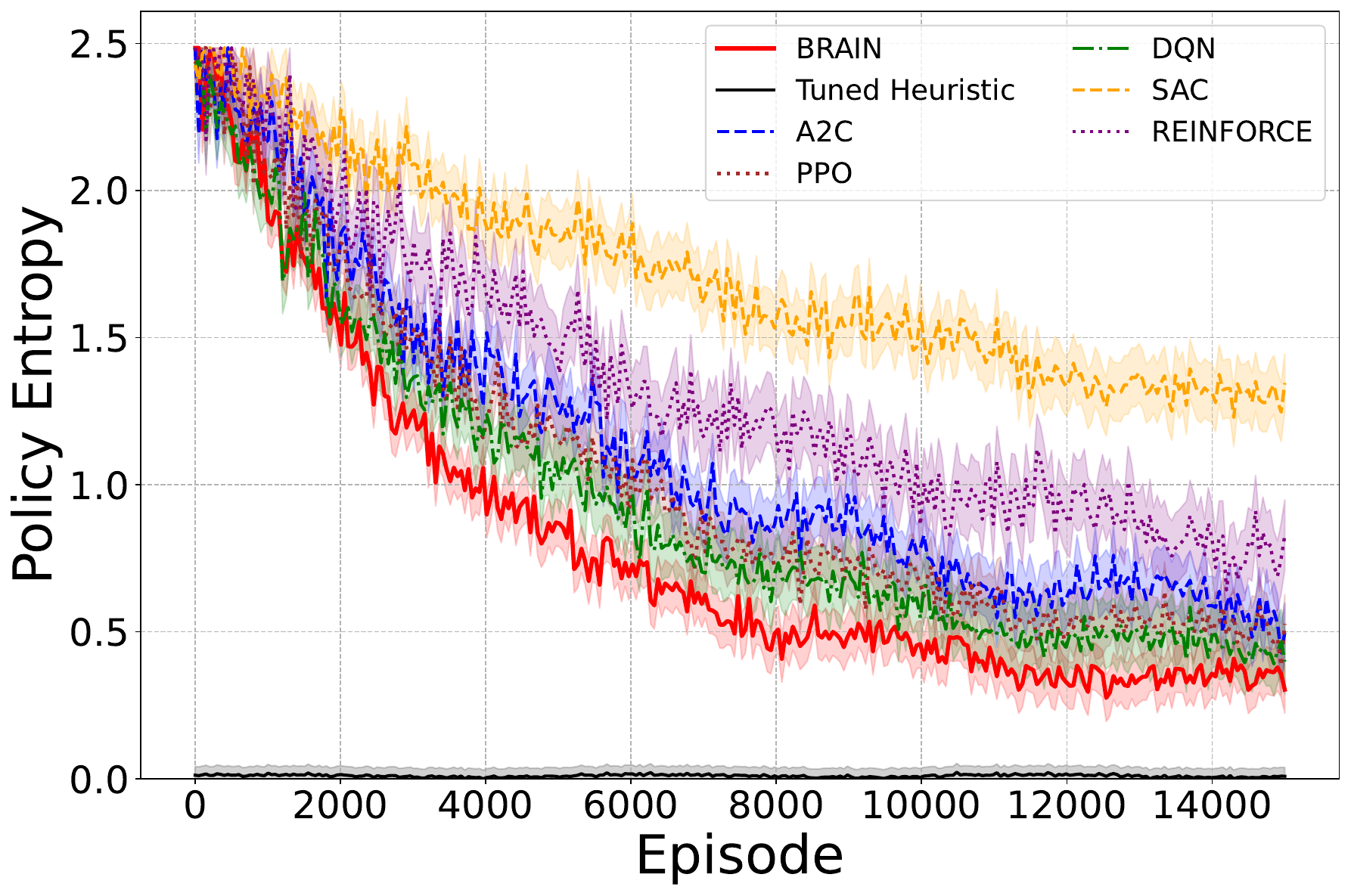}%
        \label{fig_entropy}%
    }%
    \caption{Training dynamics of agentic (RL/DRL) and embodied (active-inference) Agents on  AI-RAN testbed: \textit{i)} Mean cumulative reward per episode (higher is better), showing convergence speed and asymptotic control performance.
    \textit{ii)} Average training loss (lower is better), used as a stability proxy for optimization dynamics during online learning. \textit{iii)} Policy entropy (higher indicates more exploration), capturing the exploration-exploitation evolution over training.}
    \label{fig_baseline}
    \vspace{-.8em}
\end{figure*}

\begin{figure*}
    \centering
    \subfloat[Belief State H. - {\fontfamily{qcr}\selectfont
eMBB Slice}.]
    {%
        \includegraphics[width=0.33\linewidth]{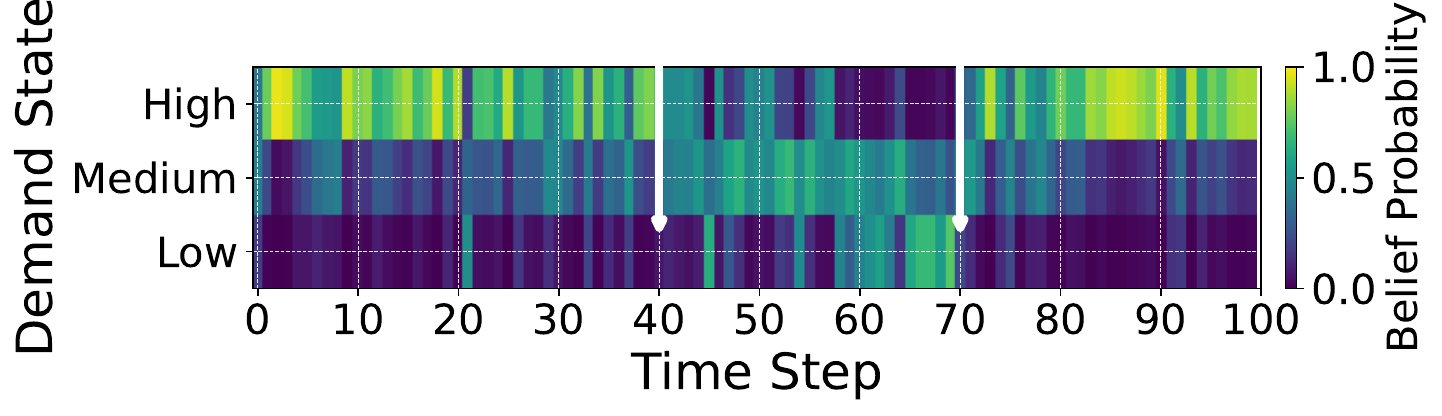}%
        \label{fig_beliefa}%
    }%
    \hfill
    \subfloat[Belief State H.- {\fontfamily{qcr}\selectfont
URLLC Slice}.]{%
        \includegraphics[width=0.33\linewidth]{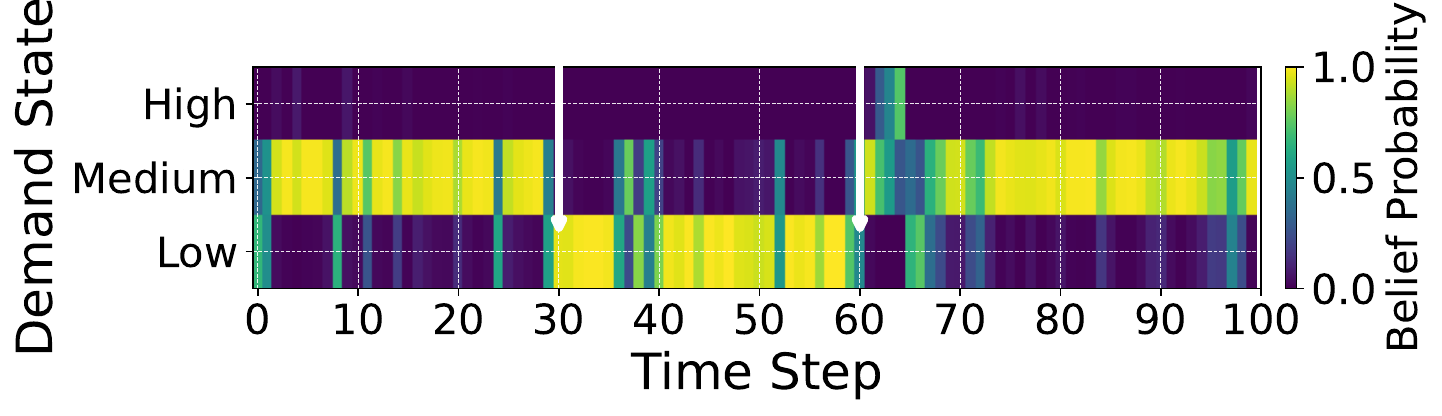}%
        \label{fig_beliefb}%
    }%
    \hfill
    \subfloat[Belief State H. - {\fontfamily{qcr}\selectfont
mMTC Slice}.]{%
        \includegraphics[width=0.33\linewidth]{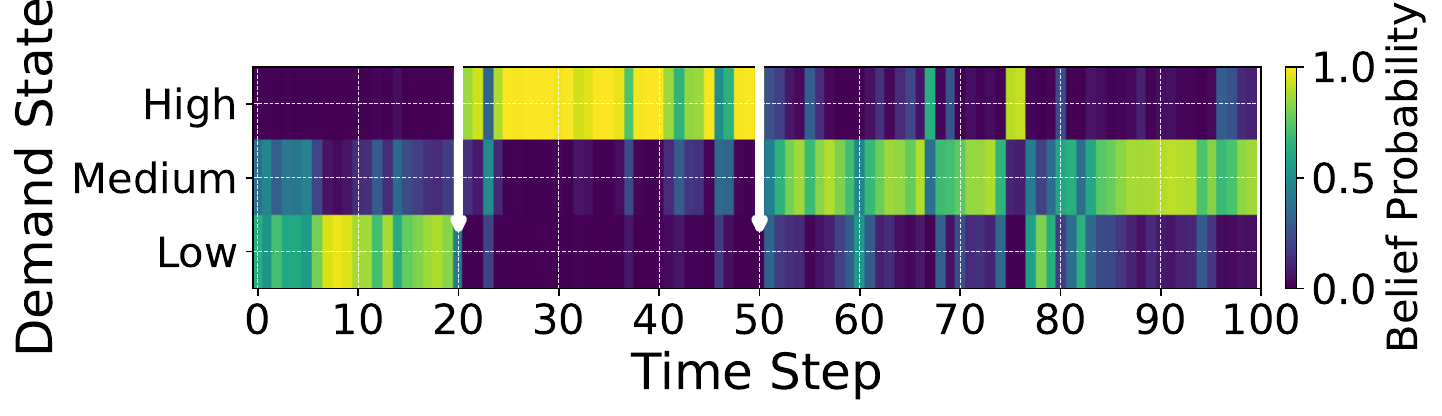}%
        \label{fig_beliefc}%
    }%
    \caption{Agent’s posterior belief trajectory over hidden traffic demand levels ({\fontfamily{qcr}\selectfont
Low}, {\fontfamily{qcr}\selectfont
Medium}, {\fontfamily{qcr}\selectfont
High}) for each network slice across episodes. Time on the x-axis indexes discrete decision epochs (e.g., observation-update intervals $\times 10^3$), and the y-axis enumerates the three demand states ({\fontfamily{qcr}\selectfont
Low} at bottom to {\fontfamily{qcr}\selectfont
High} at top). Color encodes the belief probability $Q(s_t)$ (\textit{brighter/yellow = higher, darker/purple = lower}). White arrows mark the agent’s explicit information-gathering {\fontfamily{qcr}\selectfont
Check} actions.}
    \label{fig_heatmap}
    \vspace{-.8em}
\end{figure*}

\begin{figure*}
    \centering
    \subfloat[EFE Components - {\fontfamily{qcr}\selectfont
eMBB Slice}.]
    {%
        \includegraphics[width=0.33\linewidth]{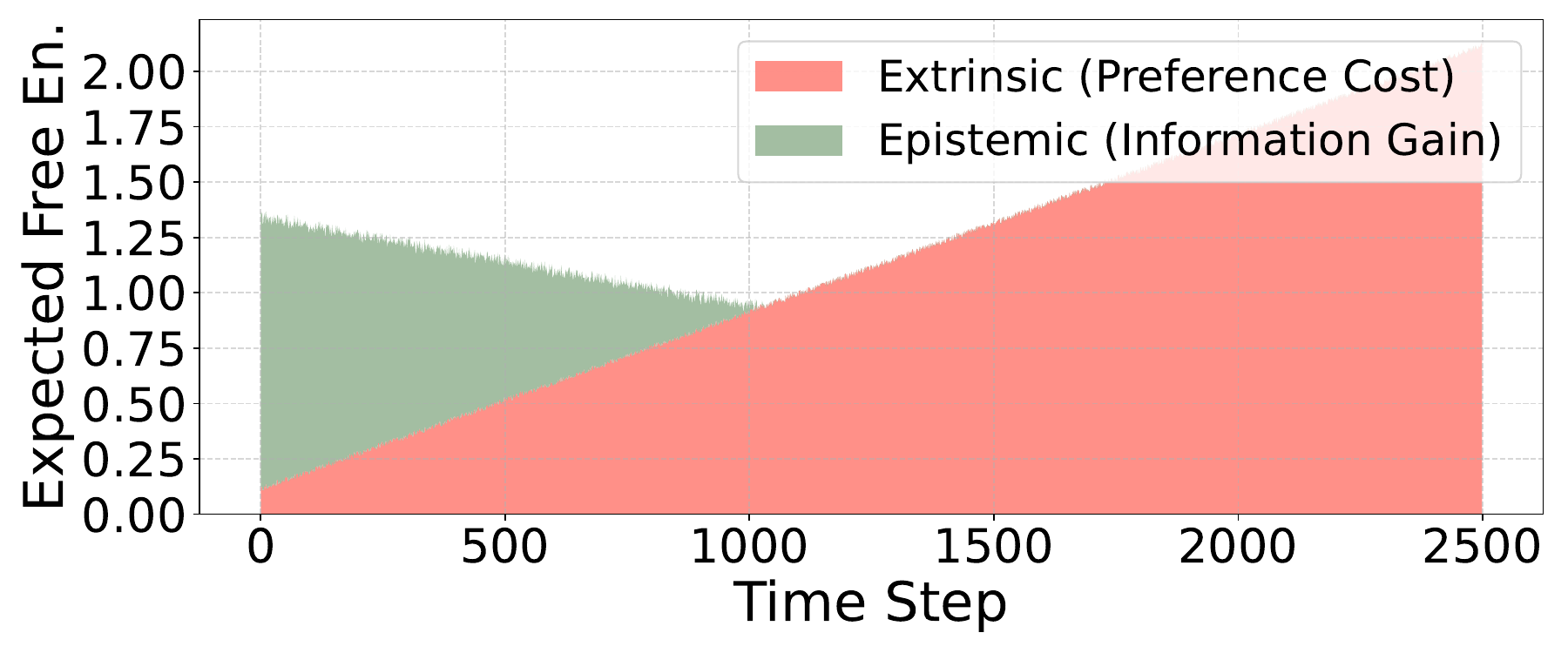}%
        \label{fig_freenergya}%
    }%
    \hfill
    \subfloat[EFE Components - {\fontfamily{qcr}\selectfont
URLLC Slice}.]{%
        \includegraphics[width=0.33\linewidth]{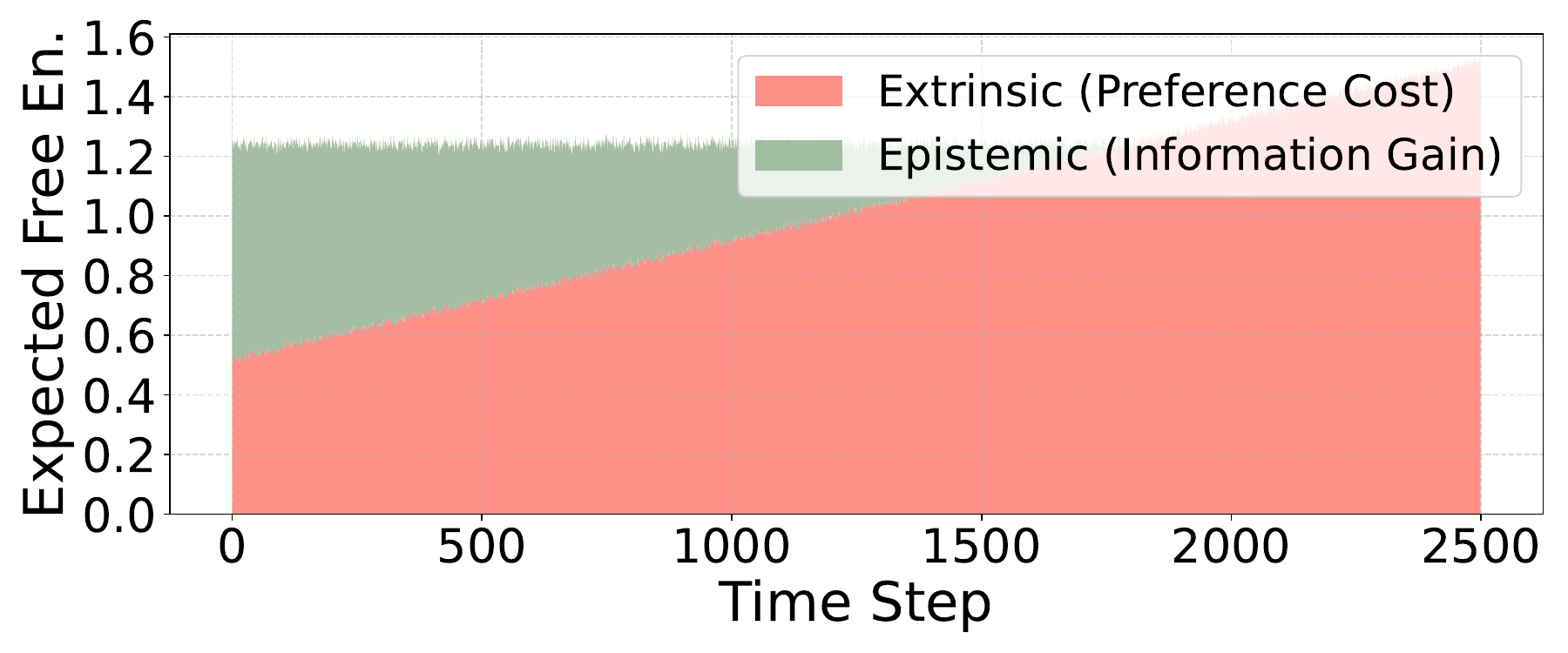}%
        \label{fig_freenergyb}%
    }%
    \hfill
    \subfloat[EFE Components - {\fontfamily{qcr}\selectfont
mMTC Slice}.]{%
        \includegraphics[width=0.33\linewidth]{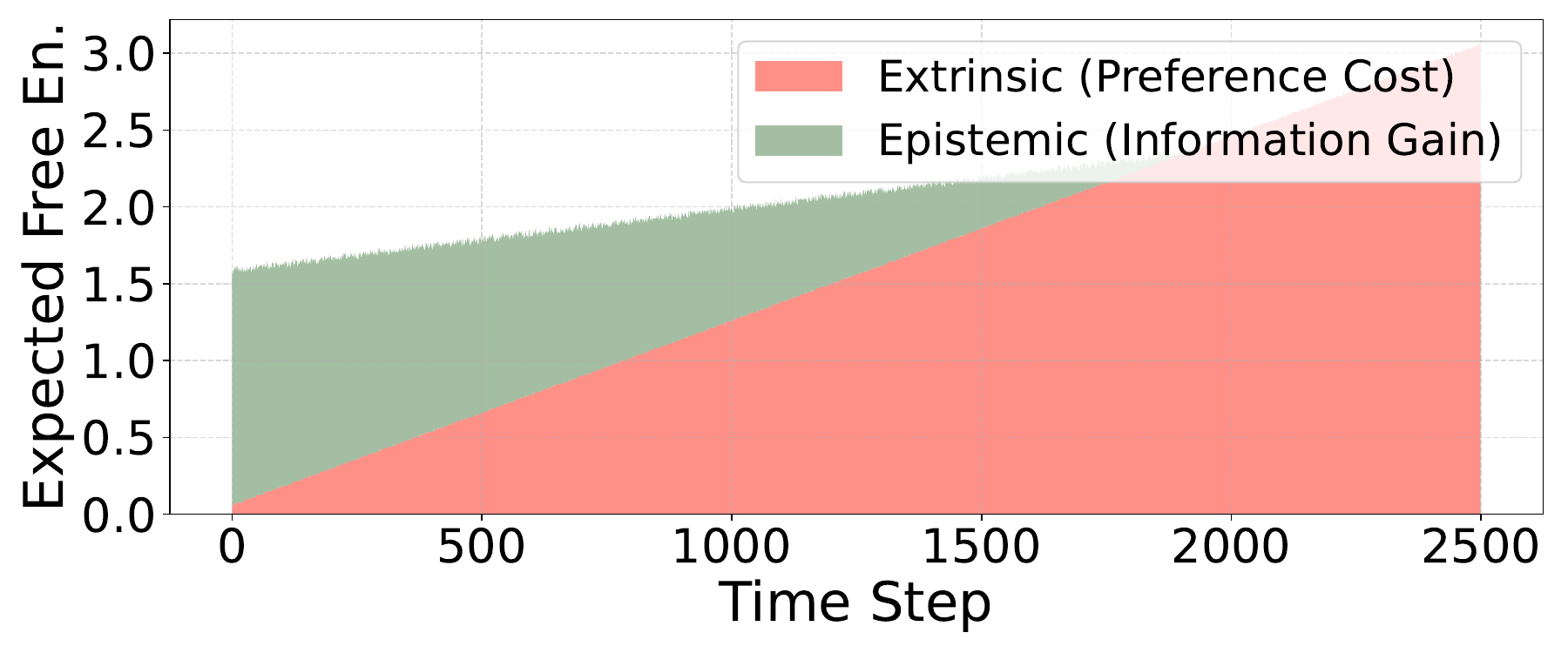}%
        \label{fig_freenergyc}%
    }%
    \caption{Interpretation of the \gls{efe} ($G(\pi))$ decomposition. At each time step, the chosen action's epistemic (soft green) and extrinsic (soft red) terms presented.  Epistemic value dominates early on (favoring exploration) and then gives way to extrinsic value (favoring QoS/exploitation).}
    \label{fig_freenergy}
    \vspace{-.8em}
\end{figure*}

\subsection{Use Case: Intelligent Orchestration}

We consider a multi-slice RAN scenario where an intelligent xApp is deployed as an autonomous agent for closed-loop slice resource orchestration. \gls{oran} near-RT RIC hosts our {\fontfamily{qcr}\selectfont BRAIN} xApp, which observes network state and dynamically controls a gNB serving multiple slices. In our setup, a single 100~MHz cell (one O-RU/gNB) serves two UEs with three slice types; \gls{embb}, \gls{urllc}, and massive \gls{mmtc} each with distinct QoS requirements (high throughput for \gls{embb}, low latency for \gls{urllc}, high reliability for \gls{mmtc}). The gNB’s MAC is slice-aware, maintaining separate buffers and scheduler queues per slice, and the UEs generate traffic for their respective slices (e.g., one UE carries a video stream on \gls{embb} and a real-time control flow on \gls{urllc} via separate PDU sessions, while the other carries sporadic IoT telemetry on \gls{mmtc}). This forms a rich environment for the \gls{xapp} agent to orchestrate intelligently.

\begin{table}
    \caption{Hyperparameters of baseline \gls{rl} models.}
    \centering
    \begin{tabularx}{\columnwidth}{l >{\centering}X >{\centering}X >{\centering\arraybackslash}X}
    \hline
    Model & Learning R. & Discount F. & Replay Buffer \\
    \hline
    Deep Q-Network (DQN) & $1\times10^{-3}$ & 0.99 & $10^5$ exp. \\
    Actor-Critic (A2C)  & $1\times10^{-3}$ & 0.99 & on-policy \\
    Policy Gradient (REINFORCE) & $5\times10^{-4}$ & 0.99 & on-policy \\
    Proximal Policy Opt. (PPO) & $5\times10^{-4}$ & 0.99 & on-policy \\
    Soft Actor-Critic (SAC) & $1\times10^{-3}$ & 0.99 & $10^5$ exp. \\
    \hline
    \end{tabularx}
    \label{baseline_conf}
    \vspace{-.8em}
\end{table}

The {\fontfamily{qcr}\selectfont BRAIN} xApp continuously monitors slice performance via the \gls{oran} E2 interface. A lightweight \gls{kpm}  monitor \gls{xapp} streams real-time telemetry \cite{near-realtime-ric-installation}; such as slice-specific downlink throughput, buffer occupancy (queue length), and downlink transport block count into the RIC’s data layer (using the standard \gls{oran} \gls{kpm} service model). These metrics, shown in prior work \cite{polese2023coloran} to effectively capture slice traffic demand and \gls{qos} conditions, constitute the state ${s}$ that our agent observes. At each control interval (on the order of tens of milliseconds), {\fontfamily{qcr}\selectfont BRAIN} computes an action ${a}$ to adjust the RAN slicing policy. The action space includes tuning the \gls{prb} allocation fraction for each slice (partitioning the cell’s bandwidth among \gls{embb}/\gls{urllc}/\gls{mmtc}) and selecting the scheduling algorithm per slice (e.g., proportional fair, round-robin, or weighted fair queueing). These commands are dispatched to the gNB via an \gls{oran} E2 control message (using a custom control service model aligning with \gls{oran} specifications), thereby closing the control loop. In this agentic deployment, the xApp autonomously adapts network parameters in real time to satisfy slice service-level objectives.

Slice-specific \gls{qos} priorities are encoded into the agent’s reward (or utility) function to drive its behavior. In our design, the \gls{embb} slice is throughput-oriented (the agent rewards high \gls{embb} data rates), the \gls{urllc} slice is latency-sensitive (the agent penalizes large \gls{urllc} buffer occupancy to minimize queuing delay), and the \gls{mmtc} slice is reliability-focused (the agent rewards successful delivery of \gls{mmtc} transport blocks, which correlates with reliable coverage for sporadic IoT traffic). Guided by these objectives, the {\fontfamily{qcr}\selectfont BRAIN} \gls{xapp} can, for example, allocate extra \gls{prb}s to \gls{urllc} during congestion to promptly drain its queue, or switch the \gls{mmtc} slice’s scheduler to a more opportunistic mode when sporadic uplink packets arrive. The constantly updated \gls{kpm} state from the monitor \gls{xapp} allows {\fontfamily{qcr}\selectfont BRAIN} to verify if slice performance indicators are being met and react quickly if not.

\subsection{Baseline Agents and Training Methodology} \label{benchmarks}

\textbf{Tuning Baseline Agents.} To evaluate our {\fontfamily{qcr}\selectfont BRAIN} agent, we compare it against a broader suite of baseline agents, including both learned policies and a heuristic scheduler. Specifically, we implement: \ballnumber{1}  \textit{Tuned Heuristic} that statically partitions \gls{prb}s among slices (according to fixed priority weights) and uses a weighted round-robin scheduler (a non-learning baseline), \ballnumber{2} Deep Q-Network (DQN) agent for slice control \cite{mnih2013playing}, \ballnumber{3} Advantage Actor-Critic (A2C) agent \cite{van-hasselt2016deep}, \ballnumber{4} Vanilla Policy Gradient agent (REINFORCE \cite{williams1992simple}), \ballnumber{5} Proximal Policy Optimization (PPO) agent \cite{schulman2017proximal}, and \ballnumber{6} Soft Actor-Critic (SAC) agent \cite{haarnoja2018soft}. All learning-based \gls{xapp}s observe the same state (slice \gls{kpm} metrics) and produce the same type of actions (PRB allocations and scheduler choices) as {\fontfamily{qcr}\selectfont BRAIN}, ensuring a fair comparison. For fairness, we also give each RL agent a comparable model architecture (a 5-layer fully-connected neural network with $\sim$30 neurons per layer) and tune their hyperparameters accordingly (see \cref{baseline_conf}). In particular, all use a discount factor $\gamma=0.99$ and learning rates on the order of $10^{-3}$, with on-policy methods (A2C, REINFORCE, PPO) relying on fresh trajectory rollouts and off-policy methods (DQN, SAC) utilizing experience replay buffers. All custom agents are implemented in PyTorch and deployed as containerized \gls{xapp}s, making them compatible with the \gls{oran} near-RT RIC platform. We train each agent through direct interaction with our RAN testbed in an online learning fashion (agentic \gls{xapp}s continuously updates its policy as it receives new observations and rewards). To ensure a fair evaluation, every learning agent is trained for the same number of time steps (on the order of $10^5$ environment interactions, which equates to several hours of network time at a 20~ms control interval). We repeat each training experiment across multiple random seeds (e.g., 5 independent runs per agent) and report the mean performance with 95\% confidence intervals to account for stochastic variability. We compare both convergence speed in terms of training iterations and actual wall-clock time, since on-policy methods like PPO require more interactions but less computation per step, whereas off-policy methods like SAC can learn from fewer interactions at the cost of more intensive updates. 

\begin{figure*}
    \centering
    \subfloat[{\fontfamily{qcr}\selectfont
eMBB Slice}.]{%
        \includegraphics[width=0.30\linewidth]{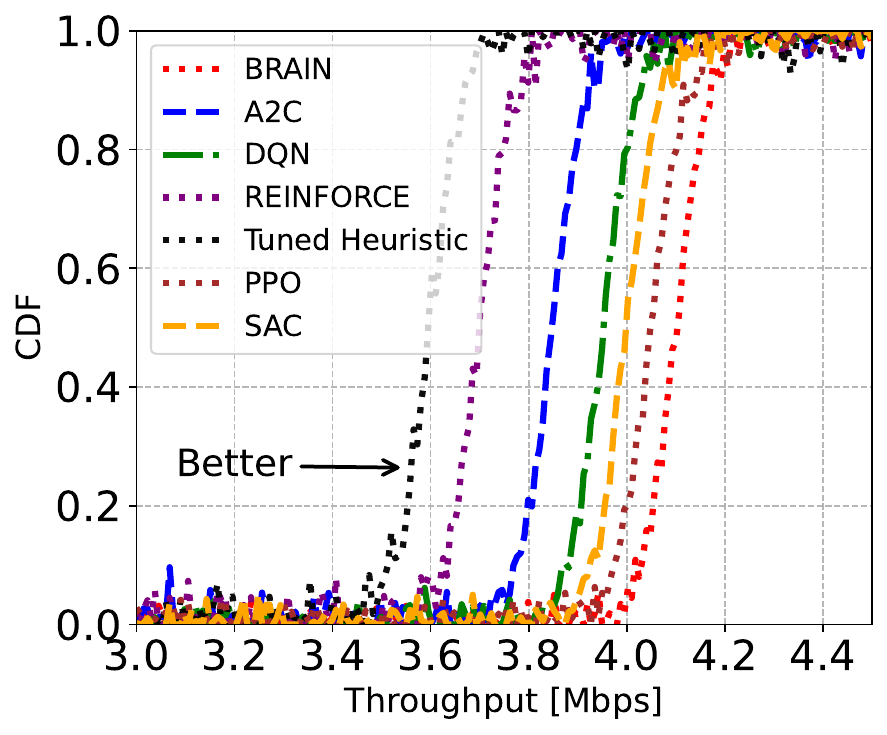}%
        \label{embb}%
    }%
    \hfill
    \subfloat[{\fontfamily{qcr}\selectfont
URLLC Slice}.]{%
        \includegraphics[width=0.30\linewidth]{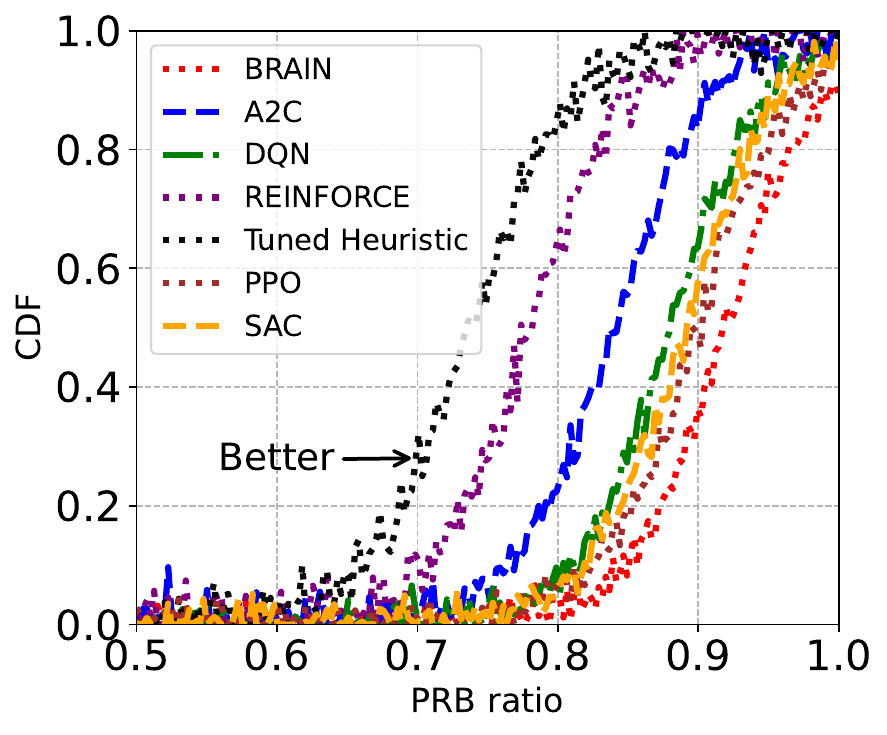}%
        \label{urllcprb}%
    }%
    \hfill
    \subfloat[{\fontfamily{qcr}\selectfont
mMTC Slice}.]{%
        \includegraphics[width=0.30\linewidth]{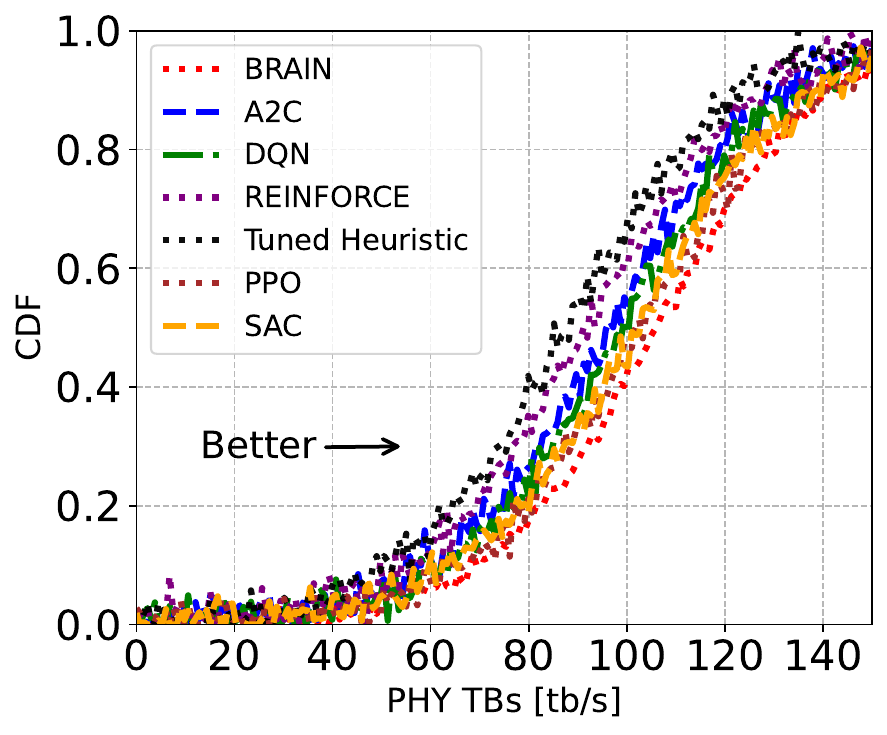}%
        \label{mtcresult}%
    }%
    \caption{Comparative network slicing performance of the proposed {\fontfamily{qcr}\selectfont BRAIN} agent against a broader set of baselines: Tuned Heuristic (weighted round-robin), and DRL agents DQN, A2C (AC), REINFORCE (PG), PPO, and SAC, evaluated across three heterogeneous network slices under the same state/action interface for a fair comparison.}
    \label{fig_perrr}
    \vspace{-.8em}
\end{figure*}

\begin{figure}
    \centering
        \includegraphics[width=0.85\linewidth]{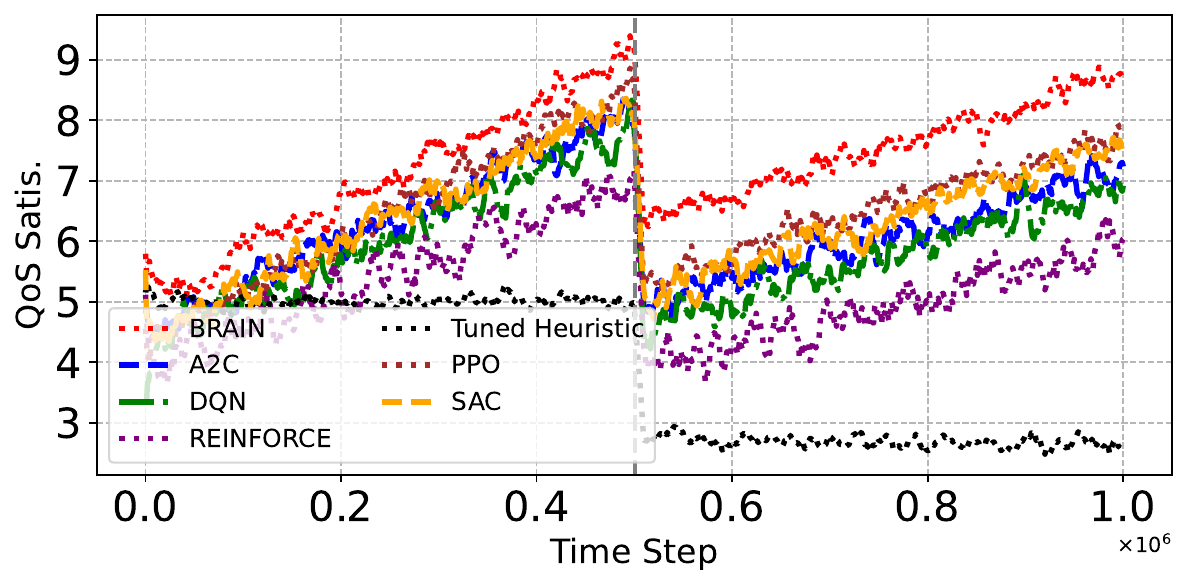}%
        \label{fig_non1}%
    
    \caption{Resilience to a controlled non-stationarity event (traffic distribution shift) measured via all-slices QoS satisfaction. The vertical dashed line at time step $0.5x10^6$ marks the non-stationary event traffic surge. }
    \label{fig_nonstation}
    \vspace{-.8em}
\end{figure}

\textbf{Tranining.} To characterize the \textit{exploration-exploitation} behavior of all controllers with a common scalar, we report \emph{policy entropy} over training time. For each episode, we compute the Shannon entropy of the \emph{action-selection distribution} used to generate decisions at each time step and average it across the episode:
\begin{align}
H_t &= -\sum_{a \in \mathcal{A}} p_t(a)\,\log\!\big(p_t(a)\big), \label{eq:entropy_step}\\
\bar{H}_{\text{ep}} &= \frac{1}{T}\sum_{t=1}^{T} H_t, \label{eq:entropy_episode}
\end{align}
where $\mathcal{A}$ denotes the (discrete) action space and $p_t(a)$ is the probability of selecting action $a$ at time step $t$. For comparability across methods, we compute entropy over the same action abstraction used by the xApps, i.e., the joint decision over PRB allocation templates and scheduler choices.

For or on-policy methods, the action-selection distribution is the learned stochastic policy, hence $p_t(a)=\pi_{\theta}(a\mid s_t)$. We compute $H_t$ directly from the categorical action probabilities output by the actor network at each time step and then average over the rollout trajectory within the episode using~\eqref{eq:entropy_episode}. This measure is consistent with the entropy quantities commonly used for entropy regularization in on-policy RL, and it captures how quickly the learned policy collapses from broad exploration (high entropy) to confident decisions (lower entropy) during training. For SAC, the action-selection distribution is also the learned stochastic actor, i.e., $p_t(a)=\pi_{\theta}(a\mid s_t)$. We compute entropy from the actor distribution at each time step and average per episode. Because SAC explicitly optimizes an entropy-regularized objective, its entropy often remains higher for longer than standard on-policy methods, reflecting sustained stochasticity in the learned control policy even after rewards stabilize.

DQN does not parameterize an explicit stochastic policy; it learns action-values $Q_{\theta}(s,a)$. To enable a meaningful entropy comparison, we define $p_t(a)$ as the \emph{behavior policy} used to select actions during training. With $\epsilon$-greedy exploration, this distribution is:
\begin{equation}
p_t(a)=\frac{\epsilon_t}{|\mathcal{A}|} + (1-\epsilon_t)\,\mathbb{I}\!\left[a=\arg\max_{a'} Q_{\theta}(s_t,a')\right], \label{eq:egreedy}
\end{equation}
and the corresponding entropy is computed via~\eqref{eq:entropy_step}--\eqref{eq:entropy_episode}. This yields an episode-wise entropy curve that reflects the exploration schedule (e.g., annealing $\epsilon_t$) and the induced concentration of behavior on the greedy action over time. The tuned heuristic applies a deterministic rule (static PRB partitioning with weighted round-robin scheduling). Consequently, the induced action-selection distribution is approximately a point mass and its entropy is near zero, serving as a reference that highlights the absence of learned exploration in the non-learning baseline.

For {\fontfamily{qcr}\selectfont BRAIN}, we define policy entropy using the posterior distribution over actions inferred from \gls{efe}. At each time step, {\fontfamily{qcr}\selectfont BRAIN} evaluates the \gls{efe} $G(a\mid s_t)$ for each candidate action and forms an action posterior:
\begin{equation}
q_t(a)=\mathrm{softmax}\!\left(-\gamma\,G(a\mid s_t)\right)
= \frac{\exp\!\left(-\gamma\,G(a\mid s_t)\right)}{\sum_{a'\in\mathcal{A}} \exp\!\left(-\gamma\,G(a'\mid s_t)\right)}, \label{eq:brain_posterior}
\end{equation}
where $\gamma$ is a precision (inverse-temperature) parameter controlling decisiveness. We then set $p_t(a)=q_t(a)$ and compute $H_t$ and $\bar{H}_{\text{ep}}$ using~\eqref{eq:entropy_step}-\eqref{eq:entropy_episode}. This entropy has a principled interpretation in active inference: it quantifies uncertainty in the inferred control decision given the agent's beliefs and preferences, and it typically decreases as the posterior sharpens with experience (without necessarily collapsing to zero if stochasticity is retained for robustness). In addition to overall performance, we design experiments to probe each controller’s adaptability and robustness. We introduce controlled non-stationary during training/deployment. For instance, a sudden change in traffic intensity or a switch in channel conditions partway through an experiment to evaluate how quickly each agent re-adjusts to new network dynamics. This tests for resilience to changing conditions and potential \textit{catastrophic forgetting} in the RL baselines (i.e., whether a policy trained under one traffic profile fails when the profile shifts). We also conduct sensitivity analyses on key parameters of our {\fontfamily{qcr}\selectfont BRAIN} agent, including the slice preference model and reward weighting (extrinsic \gls{qos} objectives vs. epistemic exploration bonus), the planning horizon length used in its decision-making, and the level of observation noise in the state input. By varying these factors, we assess how robust the agent’s performance is to mis-specified preferences or uncertainty.

\section{Evaluation}

\textbf{Analyzing Performance of Intelligent Agents.} \cref{fig_baseline} summarizes the training performance of the {\fontfamily{qcr}\selectfont BRAIN} agent versus a tuned heuristic baseline and various \gls{drl} agents (A2C, PPO, DQN, SAC, REINFORCE) in AI-RAN testbed. In \cref{fig_reward},  {\fontfamily{qcr}\selectfont BRAIN} agent’s reward curve climbs steeply, converging in far fewer episodes and reaching a higher asymptotic reward than all baselines (including the tuned heuristic). This indicates that {\fontfamily{qcr}\selectfont BRAIN} learns an effective policy with significantly higher sample efficiency; extracting more cumulative reward from limited interactions. By contrast, the \gls{drl} agents exhibit slower reward growth and lower plateaus, reflecting the heavy trial-and-error search typical of model-free \gls{rl}. Faster reward convergence means BRAIN can attain near-optimal control decisions with much less training data than the DRL benchmarks which is a critical advantage in real-world networks where each training episode (e.g. a timeslot of suboptimal decisions) has tangible costs. 
\cref{fig_loss} plots the average training loss, where again {\fontfamily{qcr}\selectfont BRAIN} distinguishes itself with a markedly lower and more stable loss trajectory throughout training.  {\fontfamily{qcr}\selectfont BRAIN} agent’s loss remains near an order of magnitude lower than the deep RL agents’ losses and shows minimal oscillation. This stability indicates that {\fontfamily{qcr}\selectfont BRAIN}’s learning updates are well-behaved, preventing the large gradient swings or divergence issues that often plague \gls{drl} training. In contrast, the RL baselines (especially more volatile ones like DQN or REINFORCE) exhibit higher loss values with noticeable fluctuations, signaling less stable learning. Such instability in \gls{rl} can arise from the algorithms struggling to adjust to the RAN’s non-stationary dynamics: when the environment’s “rules” (e.g. user load, channel conditions) change continually, a conventional RL agent has trouble re-using prior knowledge and may need to relearn repeatedly.
\cref{fig_baseline} illustrates the policy entropy over time, shedding light on each agent’s exploration–exploitation balance. {\fontfamily{qcr}\selectfont BRAIN}’s entropy starts high (encouraging exploration) and then declines gradually as training progresses. Importantly, it never collapses to zero; instead, {\fontfamily{qcr}\selectfont BRAIN}’s entropy tapers to a moderate level, indicating a controlled exploration strategy. This steady entropy reduction suggests {\fontfamily{qcr}\selectfont BRAIN}  is systematically exploring the action space early on and then confidently exploiting its learned policy as it converges, all without prematurely losing diversity in its decisions.

\textbf{Explainability Analysis.}
We model each slice’s demand as a hidden state ({\fontfamily{qcr}\selectfont Low}/{\fontfamily{qcr}\selectfont Medium}/{\fontfamily{qcr}\selectfont High}) and visualize the agent’s posterior belief over time as heatmaps in \cref{fig_heatmap}.
In the {\fontfamily{qcr}\selectfont eMBB Slice} (\cref{fig_beliefa}), the agent quickly concentrates its belief on the {\fontfamily{qcr}\selectfont High} demand state (bright band in the top row) once high traffic is observed.
Around $t\approx40$, the underlying demand drops and the belief becomes less concentrated, spreading across states before refocusing on {\fontfamily{qcr}\selectfont Medium}.
After the arrow-labeled {\fontfamily{qcr}\selectfont Check} actions, the belief sharpens again and returns to {\fontfamily{qcr}\selectfont High} by $t\approx70$.
This matches the intuition that the agent becomes more certain when informative observations arrive (a single dominant bright band) and remains more uncertain otherwise (belief distributed across multiple states). In \cref{fig_beliefb}, the posterior initially alternates mainly between {\fontfamily{qcr}\selectfont High} and {\fontfamily{qcr}\selectfont Low}, reflecting ambiguity in early observations.
At $t\approx30$, a demand shift drives rapid belief concentration toward {\fontfamily{qcr}\selectfont Low}.
Following the epistemic {\fontfamily{qcr}\selectfont Check} at $t\approx60$, the belief becomes more peaked and consolidates strongly on {\fontfamily{qcr}\selectfont Medium} demand. In \cref{fig_beliefc}, the agent begins with low uncertainty, confidently focusing on {\fontfamily{qcr}\selectfont Low}.
After {\fontfamily{qcr}\selectfont Check} at $t\approx20$, the belief shifts decisively toward {\fontfamily{qcr}\selectfont High}.
Around $t\approx50$, changing conditions increase uncertainty (belief spreads across states) before gradually concentrating toward the {\fontfamily{qcr}\selectfont Medium} state by $t\approx70$.

In \cref{fig_freenergya} for the {\fontfamily{qcr}\selectfont
eMBB Slice}, we observe that the epistemic value dominates in the early phase, where the green area is most prominent.
This indicates that the agent is initially exploring uncertain aspects of {\fontfamily{qcr}\selectfont eMBB} traffic demands, likely performing observation-driven or probing actions to refine its internal beliefs about bandwidth requirements.
Over time, the epistemic term steadily declines, while the extrinsic cost increases.
This transition reflects that the agent has gained enough confidence in its beliefs and begins to shift toward exploitative behavior, focusing on aligning slice resource allocations with performance preferences. 
In \cref{fig_freenergyb} for the {\fontfamily{qcr}\selectfont URLLC Slice}, a slightly different pattern emerges.
The epistemic and extrinsic components are more balanced during the early stages, implying that the agent simultaneously explores and regulates {\fontfamily{qcr}\selectfont URLLC}'s latency-critical requirements.
This behavior reflects the tight \gls{qos} constraints of {\fontfamily{qcr}\selectfont URLLC}, which necessitate that even early decisions consider extrinsic risks.
In \cref{fig_freenergyc} for {\fontfamily{qcr}\selectfont mMTC Slice}, we see the strongest and longest-lasting epistemic engagement.
The green region dominates the first half of the plot, suggesting that the agent initially dedicates extensive exploration effort to understand {\fontfamily{qcr}\selectfont mMTC}'s demand dynamics, which are likely bursty and sparse in nature.
After $t=2000$, a sharp increase in extrinsic value occurs as the agent begins enforcing goal-directed behavior.

\textbf{Slicing Performance.} \cref{fig_perrr} reports \emph{per-slice empirical CDFs} of \gls{kpm}s for the three heterogeneous slices, measured on the AI-RAN testbed under the same state/action interface for all agents. 
Using CDFs (rather than only means) is important because it exposes \emph{tail behavior} and reliability: a right-shift of the CDF indicates that an agent achieves larger \gls{kpm} values more frequently {\fontfamily{qcr}\selectfont
("Better $\rightarrow$")}, while a steeper CDF indicates reduced variability (more predictable operation). \cref{embb} shows that {\fontfamily{qcr}\selectfont BRAIN} yields the most favorable throughput distribution; relative to all \gls{drl} baselines and the tuned heuristic, indicating higher throughput across essentially the entire operating range.
Qualitatively, {\fontfamily{qcr}\selectfont BRAIN} improves not only the median throughput but also the upper quantiles, suggesting that the agent learns a slicing policy that preserves \gls{embb} capacity even while meeting stricter \gls{urllc}/\gls{mmtc} requirements.
In contrast, baselines exhibit either \textit{i)} lower medians or \textit{ii)} larger dispersion, implying less consistent eMBB service under the same traffic mix and control budget.
\cref{urllcprb} reports the distribution of the \gls{urllc} \gls{prb} ratio (i.e., the fraction of physical resources effectively assigned/available to \gls{urllc} by the slicing and scheduling decisions).
Higher \gls{urllc} PRB-ratio CDF reflects stronger \emph{resource protection} for \gls{urllc}, which is consistent with meeting latency-sensitive objectives under congestion.
{\fontfamily{qcr}\selectfont BRAIN} exhibits the most right-shifted curve, indicating that it allocates/maintains higher \gls{urllc} resource shares more reliably when needed.
This behavior aligns with the embodied active-inference design: the agent’s action posterior (formed via \gls{efe}) naturally increases precision toward \gls{urllc}-protective actions when beliefs indicate rising queue pressure, rather than relying on brittle reward shaping or episodic retraining.
Several \gls{drl} baselines (notably REINFORCE and the tuned heuristic) show substantially more mass at lower \gls{prb} ratios, which typically corresponds to periods where \gls{urllc} is under-provisioned and thus more vulnerable to queue build-up and latency violations. \cref{mtcresult} compares the distribution of delivered downlink PHY TBs for the \gls{mmtc} slice, which we use as a reliability-oriented proxy in our setup (successful TB deliveries reflect sustained service for sporadic IoT/telemetry traffic).
{\fontfamily{qcr}\selectfont BRAIN} provides a modest but consistent for the TB distribution versus \gls{drl} baselines, suggesting improved reliability \emph{without} sacrificing \gls{embb} throughput or \gls{urllc} protection.
Importantly, the low-performance tail is reduced: {\fontfamily{qcr}\selectfont BRAIN} yields fewer ''near-starvation'' intervals (very low TB rates), which is critical for \gls{mmtc} where sporadic bursts must still be delivered predictably.

Beyond average reward, we evaluate whether controllers \emph{maintain slice-specific service guarantees} under distribution shifts. Concretely, we measure how reliably each agent keeps all slices within their \gls{qos} targets before and after a controlled non-stationarity event. In \cref{fig_nonstation}, before the non-stationarity event ($k_{\text{shift}}\!\approx\!0.5\times10^6$), {\fontfamily{qcr}\selectfont BRAIN} achieves the highest all-slices QoS satisfaction, indicating it most consistently keeps \emph{all} slice constraints within target under the nominal regime. 
At $k_{\text{shift}}$, all learning-based agents exhibit an abrupt drop in $\mathrm{QoS\_Sat}(t)$ due to the traffic surge; however, {\fontfamily{qcr}\selectfont BRAIN} shows the \emph{smallest degradation} and the \emph{fastest recovery} toward its pre-shift level. In contrast, \gls{drl} baselines suffer a larger post-shift drop and recover more slowly, stabilizing at lower QoS-satisfaction levels; consistent with reduced adaptability and partial forgetting under distribution shift. 
The tuned heuristic remains largely flat and well below the learned agents throughout, confirming that static slicing policies cannot react to sudden regime changes.

\section{Conclusion}

This work demonstrates that \emph{deep active inference} is not only a conceptual fit for agentic and embodied intelligence in mobile networks, but also a practical control paradigm on a real open AI-RAN stack. We introduced {\fontfamily{qcr}\selectfont BRAIN} as an \gls{xapp} that closes the network action--perception loop through two tightly coupled operations: (i) \emph{Bayesian belief updating} over latent slice conditions from streaming \gls{kpm}s, and (ii) \emph{expected free-energy minimization} to select resource-allocation actions that jointly satisfy slice intents and reduce uncertainty. Across a GPU-accelerated AI-RAN testbed with heterogeneous  slices, {\fontfamily{qcr}\selectfont BRAIN} yielded three concrete outcomes. First, it achieved stronger slicing performance than a tuned heuristic and a broad set of DRL baselines. Second, it provided robust adaptation under non-stationarity: when the traffic distribution shifted abruptly, {\fontfamily{qcr}\selectfont BRAIN} exhibited the smallest QoS-satisfaction degradation and the fastest recovery without retraining. Third, it delivered \emph{operator-facing interpretability at runtime}. Beyond the empirical advantage, the broader insight is that active inference enables genuinely agentic, embodied control by grounding decisions in principled Bayesian belief updating rather than reward engineering. 

For future work, promising directions include extending the framework to hierarchical, multi-timescale active inference in \gls{oran}, where near-RT xApps operate under non-RT intent and policy coordination using structured generative models. Another important research can be to scale to multi-cell and multi-agent deployments, enabling coordination among xApps under interference and mobility coupling and studying distributed belief sharing under realistic telemetry and fronthaul constraints.

\printbibliography

@techreport{albarracin2023designing,
	author = {Albarracin, Mahault and Hip{\'{o}}lito, In{\^{e}}s and Tremblay, Safae Essafi and Fox, Jason G. and Ren{\'{e}}, Gabriel and Friston, Karl J. and Ramstead, Maxwell J. D.},
	title = {{Designing explainable artificial intelligence with active inference: A framework for transparent introspection and decision-making}},
	doi = {10.48550/ARXIV.2306.04025},
	institution = {arXiv},
	month = Jun,
	type = {cs.AI},
	year = {2023},
}

@article{ali2023explainable,
	author = {Ali, Sajid and Abuhmed, Tamer and El-Sappagh, Shaker and Muhammad, Khan and Alonso-Moral, Jose M. and Confalonieri, Roberto and Guidotti, Riccardo and Del Ser, Javier and D{\'{i}}az-Rodr{\'{i}}guez, Natalia and Herrera, Francisco},
	title = {{Explainable Artificial Intelligence (XAI): What we know and what is left to attain Trustworthy Artificial Intelligence}},
	doi = {10.1016/j.inffus.2023.101805},
	journal = {Elsevier Information Fusion},
	month = Nov,
	pages = {101805},
	publisher = {Elsevier},
	volume = {99},
	year = {2023},
}

@techreport{bank2021autoencoders,
	author = {Bank, Dor and Koenigstein, Noam and Giryes, Raja},
	title = {{Autoencoders}},
	doi = {10.48550/arXiv.2003.05991},
	institution = {arXiv},
	month = Apr,
	number = {2003.05991},
	type = {cs.NI},
	year = {2021},
}

@article{bariah2024ai,
	author = {Bariah, Lina and Debbah, Merouane},
	title = {{AI Embodiment Through 6G: Shaping the Future of AGI}},
	doi = {10.1109/mwc.015.2300521},
	issn = {1558-0687},
	journal = {IEEE Wireless Communications},
	month = Oct,
	number = {5},
	pages = {174--181},
	publisher = {IEEE},
	volume = {31},
	year = {2024},
}

@inproceedings{basaran2024xainomaly,
	author = {Basaran, Osman Tugay and Dressler, Falko},
	title = {{XAInomaly: Explainable, Interpretable and Trustworthy AI for xURLLC in 6G Open-RAN}},
	booktitle = {3rd International Conference on 6G Networking (6GNet 2024)},
	address = {Paris, France},
	doi = {10.1109/6GNet63182.2024.10765734},
	month = Oct,
	pages = {93--101},
	publisher = {IEEE},
	year = {2024},
}

@article{cheng2022channel,
	author = {Cheng, Xiang and Huang, Ziwei and Bai, Lu},
	title = {{Channel Nonstationarity and Consistency for Beyond 5G and 6G: A Survey}},
	doi = {10.1109/comst.2022.3184049},
	issn = {1553-877X},
	journal = {IEEE Communications Surveys \& Tutorials},
	number = {3},
	pages = {1634--1669},
	publisher = {IEEE},
	volume = {24},
	year = {2022},
}

@inproceedings{cheng2024oranslice,
	author = {Cheng, Hai and D’Oro, Salvatore and Gangula, Rajeev and Velumani, Sakthivel and Villa, Davide and Bonati, Leonardo and Polese, Michele and Melodia, Tommaso and Arrobo, Gabriel and Maciocco, Christian},
	title = {{ORANSlice: An Open Source 5G Network Slicing Platform for O-RAN}},
	booktitle = {30th ACM International Conference on Mobile Computing and Networking (MobiCom 2024)},
	address = {Washington, D.C.},
	doi = {10.1145/3636534.3701544},
	month = Nov,
	pages = {2297--2302},
	publisher = {ACM},
	year = {2024},
}

@article{dacosta2020active,
	author = {Da Costa, Lancelot and Parr, Thomas and Sajid, Noor and Veselic, Sebastijan and Neacsu, Victorita and Friston, Karl J.},
	title = {{Active inference on discrete state-spaces: A synthesis}},
	doi = {10.1016/j.jmp.2020.102447},
	journal = {Journal of Mathematical Psychology},
	month = Dec,
	pages = {102447},
	publisher = {Elsevier},
	volume = {99},
	year = {2020},
}

@inproceedings{duttagupta2025symbxrl,
	author = {Duttagupta, Abhishek and Jabbari, MohammadErfan and Fiandrino, Claudio and Fiore, Marco and Widmer, Joerg},
	title = {{SYMBXRL: Symbolic Explainable Deep Reinforcement Learning for Mobile Networks}},
	booktitle = {44th IEEE International Conference on Computer Communications (INFOCOM 2025)},
	address = {London, United Kingdom},
	doi = {10.1109/INFOCOM55648.2025.11044492},
	isbn = {979-8-3315-4370-9},
	month = May,
	pages = {1--10},
	publisher = {IEEE},
	year = {2025},
}

@article{fiandrino2023explora,
	author = {Fiandrino, Claudio and Bonati, Leonardo and D'Oro, Salvatore and Polese, Michele and Melodia, Tommaso and Widmer, Joerg},
	title = {{EXPLORA: AI/ML EXPLainability for the Open RAN}},
	doi = {10.1145/3629141},
	issn = {2834-5509},
	journal = {Proceedings of the ACM on Networking},
	month = Nov,
	pages = {1--26},
	publisher = {ACM},
	volume = {1},
	year = {2023},
}

@inproceedings{fiandrino2024aichronolens,
	author = {Fiandrino, Claudio and G{\'{o}}mez, Eloy P{\'{e}}rez and P{\'{e}}rez, Pablo Fern{\'{a}}ndez and Mohammadalizadeh, Hossein and Fiore, Marco and Widmer, Joerg},
	title = {{AIChronoLens: Advancing Explainability for Time Series AI Forecasting in Mobile Networks}},
	booktitle = {43rd IEEE International Conference on Computer Communications (INFOCOM 2024)},
	address = {Vancouver, Canada},
	doi = {10.1109/infocom52122.2024.10621134},
	month = May,
	pages = {1521--1530},
	publisher = {IEEE},
	year = {2024},
}

@online{foxconn,
	author = {Foxconn},
	title = {{Foxconn RPQN}},
	addendum = {(accessed: 15.01.2026)},
	url = {https://fcc.report/FCC-ID/2AQ68RPQN7801/5573870.pdf},
	year = {2024},
}

@article{friston2010free-energy,
	author = {Friston, Karl J.},
	title = {{The free-energy principle: a unified brain theory?}},
	doi = {10.1038/nrn2787},
	issn = {1471-0048},
	journal = {Nature Reviews Neuroscience},
	month = Jan,
	number = {2},
	pages = {127--138},
	publisher = {Springer Science and Business Media LLC},
	volume = {11},
	year = {2010},
}

@article{friston2016active,
	author = {Friston, Karl J. and FitzGerald, Thomas and Rigoli, Francesco and Schwartenbeck, Philipp and ODoherty, John and Pezzulo, Giovanni},
	title = {{Active inference and learning}},
	doi = {10.1016/j.neubiorev.2016.06.022},
	issn = {0149-7634},
	journal = {Neuroscience \& Biobehavioral Reviews},
	month = Sep,
	pages = {862--879},
	publisher = {Elsevier},
	volume = {68},
	year = {2016},
}

@article{fujii2024realworld,
	author = {Fujii, Kentaro and Isomura, Takuya and Murata, Shingo},
	title = {{Real-World Robot Control Based on Contrastive Deep Active Inference With Demonstrations}},
	doi = {10.1109/access.2024.3477306},
	issn = {2169-3536},
	journal = {IEEE Access},
	pages = {172343--172357},
	publisher = {IEEE},
	volume = {12},
	year = {2024},
}

@article{gizzini2024towards,
	author = {Gizzini, Abdul Karim and Medjahdi, Yahia and Ghandour, Ali J. and Clavier, Laurent},
	title = {{Towards Explainable AI for Channel Estimation in Wireless Communications}},
	doi = {10.1109/tvt.2023.3345632},
	issn = {1939-9359},
	journal = {IEEE Transactions on Vehicular Technology},
	month = May,
	number = {5},
	pages = {7389--7394},
	publisher = {IEEE},
	volume = {73},
	year = {2024},
}

@article{guo2020explainable,
	author = {Guo, Weisi},
	title = {{Explainable Artificial Intelligence for 6G: Improving Trust between Human and Machine}},
	doi = {10.1109/mcom.001.2000050},
	issn = {0163-6804},
	journal = {IEEE Communications Magazine},
	month = Jun,
	number = {6},
	pages = {39--45},
	publisher = {IEEE},
	volume = {58},
	year = {2020},
}

@techreport{haarnoja2018soft,
	author = {Haarnoja, Tuomas and Zhou, Aurick and Abbeel, Pieter and Levine, Sergey},
	title = {{Soft Actor-Critic: Off-Policy Maximum Entropy Deep Reinforcement Learning with a Stochastic Actor}},
	doi = {10.48550/ARXIV.1801.01290},
	institution = {arXiv},
	month = Aug,
	number = {1801.01290},
	type = {cs.LG},
	year = {2018},
}

@article{kaltenberger2020openairinterface,
	author = {Kaltenberger, Florian and Silva, Aloizio P. and Gosain, Abhimanyu and Wang, Luhan and Nguyen, Tien-Thinh},
	title = {{OpenAirInterface: Democratizing innovation in the 5G Era}},
	doi = {10.1016/j.comnet.2020.107284},
	issn = {1389-1286},
	journal = {Elsevier Computer Networks},
	month = Jul,
	pages = {107284},
	publisher = {Elsevier},
	volume = {176},
	year = {2020},
}

@inproceedings{kelkar2021aerial,
	author = {Kelkar, Anupa and Dick, Chris},
	title = {{Aerial: A GPU Hyperconverged Platform for 5G}},
	booktitle = {ACM SIGCOMM 2021, Demo Session},
	address = {Virtual Conference},
	doi = {10.1145/3472716.3472864},
	isbn = {978-1-4503-8383-7},
	month = Aug,
	pages = {79--81},
	publisher = {ACM},
	year = {2021},
}

@inproceedings{kelkar2021nvidia,
	author = {Kelkar, Anupa and Dick, Chris},
	title = {{NVIDIA Aerial GPU Hosted AI-on-5G}},
	booktitle = {4th IEEE 5G World Forum (5GWF 2021)},
	address = {Montr{\'{e}}al, Canada},
	doi = {10.1109/5gwf52925.2021.00019},
	isbn = {978-1-6654-4308-1},
	month = Oct,
	pages = {64--69},
	publisher = {IEEE},
	year = {2021},
}

@incollection{lecun1998convolutional,
	author = {LeCun, Yann and Bengio, Yoshua},
	title = {{Convolutional Networks for Images, Speech, and Time Series}},
	editor = {Arbib, M. A.},
	booktitle = {Handbook of Brain Theory and Neural Networks},
	pages = {255--258},
	publisher = {MIT Press},
	year = {2003},
}

@inproceedings{liu2021onslicing,
	author = {Liu, Qiang and Choi, Nakjung and Han, Tao},
	title = {{OnSlicing: Online End-to-End Network Slicing with Reinforcement Learning}},
	booktitle = {17th ACM International Conference on Emerging Networking Experiments and Technologies (CoNEXT 2021)},
	address = {Virtual Conference},
	doi = {10.1145/3485983.3494850},
	month = Dec,
	pages = {141--153},
	publisher = {ACM},
	year = {2021},
}

@article{liu2025embodied,
	author = {Liu, Huaping and Guo, Di and Cangelosi, Angelo},
	title = {{Embodied Intelligence: A Synergy of Morphology, Action, Perception and Learning}},
	doi = {10.1145/3717059},
	issn = {0360-0300},
	journal = {ACM Computing Surveys},
	month = Mar,
	number = {7},
	pages = {1--36},
	publisher = {ACM},
	volume = {57},
	year = {2025},
}

@article{longo2024explainable,
	author = {Longo, Luca and Brcic, Mario and Cabitza, Federico and Choi, Jaesik and Confalonieri, Roberto and Ser, Javier Del and Guidotti, Riccardo and Hayashi, Yoichi and Herrera, Francisco and Holzinger, Andreas and Jiang, Richard and Khosravi, Hassan and Lecue, Freddy and Malgieri, Gianclaudio and P{\'{a}}ez, Andr{\'{e}}s and Samek, Wojciech and Schneider, Johannes and Speith, Timo and Stumpf, Simone},
	title = {{Explainable Artificial Intelligence (XAI) 2.0: A manifesto of open challenges and interdisciplinary research directions}},
	doi = {10.1016/j.inffus.2024.102301},
	journal = {Elsevier Information Fusion},
	month = Jun,
	pages = {102301},
	publisher = {Elsevier},
	volume = {106},
	year = {2024},
}

@inproceedings{lundberg2017unified,
	author = {Lundberg, Scott M. and Lee, Su-In},
	title = {{A Unified Approach to Interpreting Model Predictions}},
	booktitle = {31st International Conference on Neural Information Processing Systems (NIPS 2017)},
	address = {Long Beach, CA},
	isbn = {978-1-5108-6096-4},
	month = Dec,
	pages = {4768--4777},
	publisher = {Curran Associates Inc.},
	year = {2017},
}

@article{maier2026from,
	author = {Maier, Martin},
	title = {{From artificial intelligence to active inference: the key to true AI and the 6G world brain [Invited]}},
	doi = {10.1364/jocn.566810},
	journal = {Journal of Optical Communications and Networking},
	month = Jan,
	number = {1},
	pages = {A28--A43},
	publisher = {OSA},
	volume = {18},
	year = {2026},
}

@techreport{mnih2013playing,
	author = {Mnih, Volodymyr and Kavukcuoglu, Koray and Silver, David and Graves, Alex and Antonoglou, Ioannis and Wierstra, Daan and Riedmiller, Martin},
	title = {{Playing Atari with Deep Reinforcement Learning}},
	doi = {10.48550/arXiv.1312.5602},
	institution = {arXiv},
	month = Dec,
	number = {1312.5602},
	type = {cs.LG},
	year = {2013},
}

@online{near-realtime-ric-installation,
	author = {{O-RAN SC}},
	title = {{OSC Near Realtime RIC}},
	addendum = {(accessed: 15.01.2026)},
	url = {https://wiki.o-ran-sc.org/display/RICP/2022-05-24+ Release+E},
	year = {2024},
}

@online{nvidiaaerial,
	author = {NVIDIA},
	title = {{NVIDIA Aerial SDK}},
	addendum = {(accessed: 15.01.2026)},
	url = {https://developer.nvidia.com/aerial- sdk},
	year = {2024},
}

@article{oliver2022empirical,
	author = {Oliver, Guillermo and Lanillos, Pablo and Cheng, Gordon},
	title = {{An Empirical Study of Active Inference on a Humanoid Robot}},
	doi = {10.1109/tcds.2021.3049907},
	issn = {2379-8920},
	journal = {IEEE Transactions on Cognitive and Developmental Systems},
	month = Jun,
	number = {2},
	pages = {462--471},
	publisher = {IEEE},
	volume = {14},
	year = {2022},
}

@book{parr2022active,
	author = {Parr, Thomas and Pezzulo, Giovanni and Friston, Karl J.},
	title = {{Active Inference: The Free Energy Principle in Mind, Brain, and Behavior}},
	doi = {10.7551/mitpress/12441.001.0001},
	isbn = {978-0-262-36997-8},
	publisher = {The MIT Press},
	year = {2022},
}

@article{polese2023coloran,
	author = {Polese, Michele and Bonati, Leonardo and D’Oro, Salvatore and Basagni, Stefano and Melodia, Tommaso},
	title = {{ColO-RAN: Developing Machine Learning-Based xApps for Open RAN Closed-Loop Control on Programmable Experimental Platforms}},
	doi = {10.1109/tmc.2022.3188013},
	issn = {1536-1233},
	journal = {IEEE Transactions on Mobile Computing},
	month = Oct,
	number = {10},
	pages = {5787--5800},
	publisher = {IEEE},
	volume = {22},
	year = {2023},
}

@article{polese2024empowering,
	author = {Polese, Michele and Dohler, Mischa and Dressler, Falko and Erol-Kantarci, Melike and Jana, Rittwik and Knopp, Raymond and Melodia, Tommaso},
	title = {{Empowering the 6G Cellular Architecture with Open RAN}},
	doi = {10.1109/JSAC.2023.3334610},
	issn = {0733-8716},
	journal = {IEEE Journal on Selected Areas in Communications},
	month = Feb,
	number = {2},
	pages = {245--262},
	publisher = {IEEE},
	volume = {42},
	year = {2024},
}

@inproceedings{prez2025chronoprof,
	author = {P{\'{e}}rez, Pablo Fern{\'{a}}ndez and Bravo, I{\~{n}}aki and Kamath, Anirudh and Fiandrino, Claudio and Widmer, Joerg},
	title = {{ChronoProf: Profiling Time Series Forecasters and Classifiers in Mobile Networks with Explainable AI}},
	booktitle = {26th IEEE International Symposium on a World of Wireless, Mobile and Multimedia Networks (WoWMoM 2025)},
	address = {Fort Worth, TX},
	doi = {10.1109/wowmom65615.2025.00019},
	month = May,
	pages = {41--50},
	publisher = {IEEE},
	year = {2025},
}

@inproceedings{ribeiro2016why,
	author = {Ribeiro, Marco Tulio and Singh, Sameer and Guestrin, Carlos},
	title = {{{``}Why Should I Trust You?{''}: Explaining the Predictions of Any Classifier}},
	booktitle = {22nd ACM SIGKDD International Conference on Knowledge Discovery and Data Mining},
	address = {San Francisco, CA},
	doi = {10.1145/2939672.2939778},
	isbn = {978-1-4503-4232-2},
	month = Aug,
	pages = {1135 -- 1144},
	publisher = {ACM},
	year = {2016},
}

@article{saad2020vision,
	author = {Saad, Walid and Bennis, Mehdi and Chen, Mingzhe},
	title = {{A Vision of 6G Wireless Systems: Applications, Trends, Technologies, and Open Research Problems}},
	doi = {10.1109/mnet.001.1900287},
	issn = {1558-156X},
	journal = {IEEE Network},
	month = May,
	number = {3},
	pages = {134--142},
	publisher = {IEEE},
	volume = {34},
	year = {2020},
}

@article{saad2025artificial,
	author = {Saad, Walid and Hashash, Omar and Thomas, Christo Kurisummoottil and Chaccour, Christina and Debbah, Merouane and Mandayam, Narayan and Han, Zhu},
	title = {{Artificial General Intelligence (AGI)-Native Wireless Systems: A Journey Beyond 6G}},
	doi = {10.1109/jproc.2025.3526887},
	issn = {0018-9219},
	journal = {Proceedings of the IEEE},
	month = Sep,
	number = {9},
	pages = {849--887},
	publisher = {IEEE},
	volume = {113},
	year = {2025},
}

@techreport{schulman2017proximal,
	author = {Schulman, John and Wolski, Rich and Dhariwal, Prafulla and Radford, Alec and Klimov, Oleg},
	title = {{Proximal Policy Optimization Algorithms}},
	doi = {10.48550/ARXIV.1707.06347},
	institution = {arXiv},
	month = Jul,
	pages = {1--12},
	type = {cs.LG},
	year = {2017},
}

@book{sutton1998reinforcement,
	author = {Sutton, Richard S. and Barto, Andrew G.},
	title = {{Reinforcement Learning: An Introduction}},
	location = {Cambridge, MA},
	pages = {322},
	publisher = {MIT Press},
	year = {1998},
}

@article{thomas2024causal,
	author = {Thomas, Christo Kurisummoottil and Chaccour, Christina and Saad, Walid and Debbah, Merouane and Hong, Choong Seon},
	title = {{Causal Reasoning: Charting a Revolutionary Course for Next-Generation AI-Native Wireless Networks}},
	doi = {10.1109/mvt.2024.3359357},
	issn = {1556-6072},
	journal = {IEEE Vehicular Technology Magazine},
	month = Mar,
	number = {1},
	pages = {16--31},
	publisher = {IEEE},
	volume = {19},
	year = {2024},
}

@techreport{tschantz2020reinforcement,
	author = {Tschantz, Alexander and Millidge, Beren and Seth, Anil K. and Buckley, Christopher L.},
	title = {{Reinforcement Learning through Active Inference}},
	doi = {10.48550/ARXIV.2002.12636},
	institution = {arXiv},
	month = Feb,
	type = {cs.LG},
	year = {2020},
}

@inproceedings{van-hasselt2016deep,
	author = {van Hasselt, Hado and Guez, Arthur and Silver, David},
	title = {{Deep Reinforcement Learning with Double Q-learning}},
	booktitle = {30th AAAI Conference on Artificial Intelligence (AAAI 2016)},
	address = {Phoenix, AZ},
	month = Feb,
	pages = {2094--2100},
	publisher = {AAAI Press},
	year = {2016},
}

@article{vincent2018introduction,
	author = {Vincent, Fran{\c{c}}ois-Lavet and Peter, Henderson and Riashat, Islam and Marc, G. Bellemare and Joelle, Pineau},
	title = {{An Introduction to Deep Reinforcement Learning}},
	doi = {10.1561/2200000071},
	issn = {1935-8237},
	journal = {Foundations and Trends® in Machine Learning},
	month = Dec,
	number = {3--4},
	pages = {219--354},
	publisher = {Now Foundations and Trends},
	volume = {11},
	year = {2018},
}

@article{vouros2022explainable,
	author = {Vouros, George A.},
	title = {{Explainable Deep Reinforcement Learning: State of the Art and Challenges}},
	doi = {10.1145/3527448},
	issn = {0360-0300},
	journal = {ACM Computing Surveys},
	month = Dec,
	number = {5},
	pages = {1--39},
	publisher = {ACM},
	volume = {55},
	year = {2022},
}

@article{williams1992simple,
	author = {Williams, Ronald J.},
	title = {{Simple statistical gradient-following algorithms for connectionist reinforcement learning}},
	doi = {10.1007/bf00992696},
	issn = {1573-0565},
	journal = {Machine Learning},
	month = May,
	number = {3--4},
	pages = {229--256},
	publisher = {Springer Science and Business Media LLC},
	volume = {8},
	year = {1992},
}

@article{yang2025generalization,
	author = {Yang, Xincheng and Shi, Yan and Liu, Junyu and Xie, Ziwen and Sheng, Min and Li, Jiandong},
	title = {{Generalization-Enhanced DRL-Based Resource Allocation in Wireless Communication Networks With Dynamic User Loads}},
	doi = {10.1109/lwc.2025.3606974},
	issn = {2162-2337},
	journal = {IEEE Wireless Communications Letters},
	month = Dec,
	number = {12},
	pages = {3902--3906},
	publisher = {IEEE},
	volume = {14},
	year = {2025},
}


\begin{IEEEbiography}[{\includegraphics[width=1in,height=1.25in,clip,keepaspectratio]{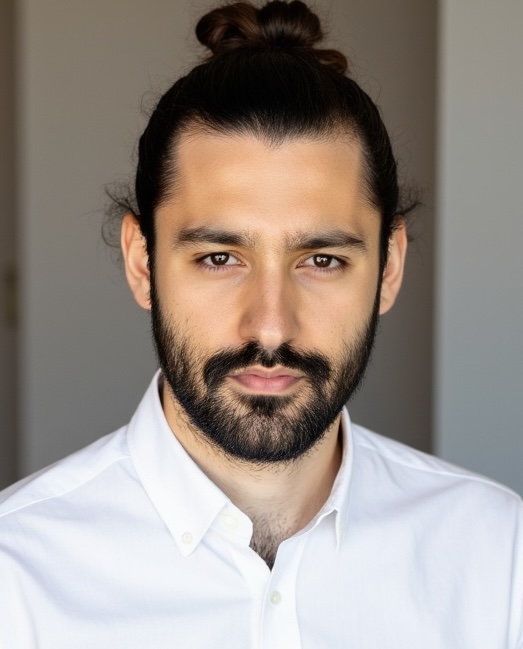}}]{Osman Tugay Basaran} is a 6G/AI Research Scientist at Prof. Dr.-Ing. Falko Dressler’s Telecommunications Networks Group (TKN), School of Electrical Engineering and Computer Science, Technical University of Berlin, Germany, while holding two Visiting Research Scientist positions at the Fraunhofer Heinrich Hertz Institute, Wireless Communications and Networks Department, Signal and Information Processing Research Group of Prof. Dr.-Ing. Slawomir
Stanczak and at Optical Zeitgeist Laboratory of Prof. Dr. Martin Maier, Montréal, Canada. Within the scope of 6G-Platform, 6G-RIC and xG-RIC projects, he is focusing on Domain-specific Generative AI (GenAI) and Explainable AI (XAI) algorithms for the implementation and execution of Next-Generation, AI-Native 6G and Beyond as well as cognitive skilled agents for future wireless systems. He has been selected as one of the Research Fellow (36 Fellow selected across all German Technical Universities) in the Software Campus Executive Leadership Cohort 2025. As a Principal Investigator in collaboration with Huawei, he is leading collaborative project titled “NEXT-G: Explainable and Trustworthy AI/ML for 6G and Beyond”.  He also serves as a Scientific Reviewer in top-tier IEEE Journal/Conferences such as different Transactions, Magazines, INFOCOM, ICC, and GLOBECOM.

\end{IEEEbiography}

\begin{IEEEbiography}[{\includegraphics[width=1in,height=1.25in,clip,keepaspectratio]{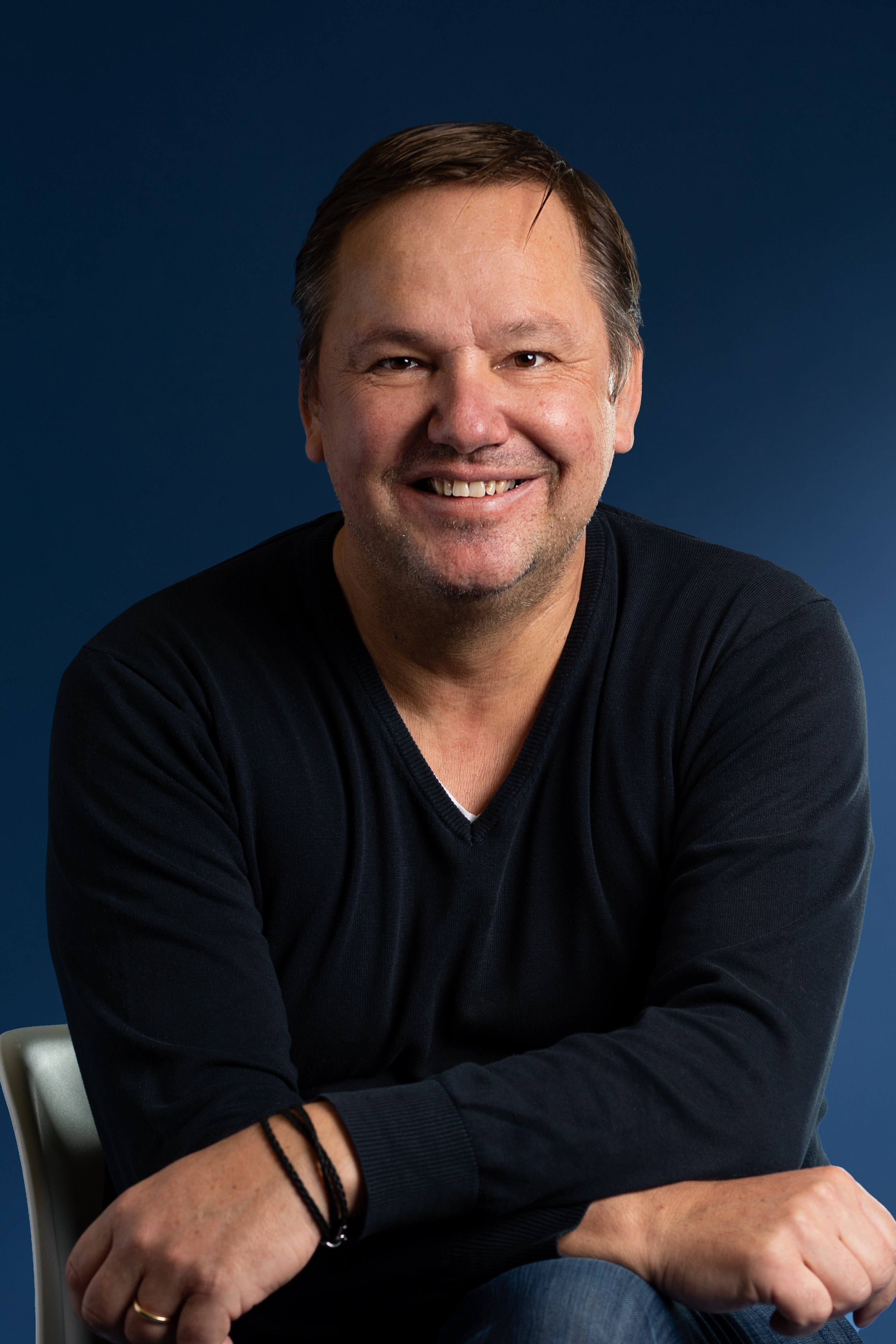}}]{Martin Maier} is a full professor with INRS, Montréal, Canada. He was educated at the Technical University of Berlin, Germany, and received M.Sc. and Ph.D. degrees both with distinctions (summa cum laude) in 1998 and 2003, respectively. In 2003, he was a postdoc fellow at the Massachusetts
Institute of Technology (MIT), Cambridge, MA. He was a visiting professor at Stanford University, Stanford, CA, 2006 through 2007. In 2017, he received the Friedrich Wilhelm Bessel Research Award from the Alexander von Humboldt (AvH) Foundation in recognition of his accomplishments in research on FiWi-enhanced mobile networks. In 2017, he was named one of the three most promising scientists in the category “Contribution to a better society” of the Marie Sklodowska-Curie Actions (MSCA) 2017 Prize Award of the European Commission. In 2019/2020, he held a UC3MBanco de Santander Excellence Chair at Universidad Carlos III de Madrid (UC3M), Madrid, Spain. Recently, in December 2023, he was awarded with the 2023 Technical Achievement Award of the IEEE Communications Society (ComSoc) Tactile Internet Technical Committee for his contribution on 6G/Next G and the design of Metaverse concepts and architectures as well as the 2023 Outstanding Paper Award of the IEEE Computer Society Bio-Inspired Computing STC for his contribution on the symbiosis between INTERnet and Human BEING (INTERBEING). He is co-author of the book “Toward 6G: A New Era of Convergence” (Wiley-IEEE Press, January 2021) and author of the sequel “6G and Onward to Next G: The Road to the Multiverse” (Wiley-IEEE Press, February 2023).
\end{IEEEbiography}

\begin{IEEEbiography}[{\includegraphics[width=1in,height=1.25in,clip,keepaspectratio]{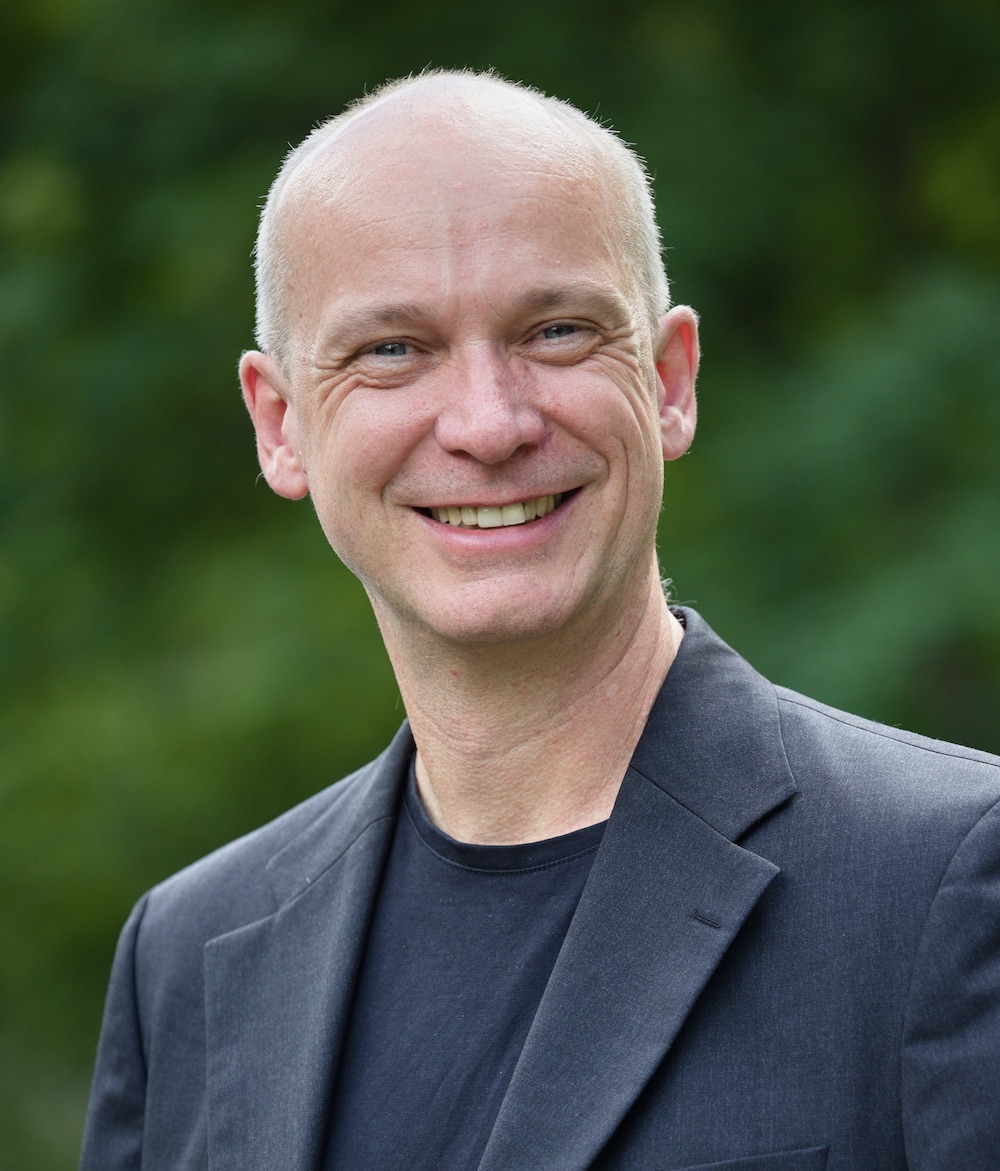}}]{Falko Dressler} is full professor and Chair for
Telecommunication Networks at the School of Electrical Engineering and Computer Science, TU Berlin. He received his M.Sc. and Ph.D. degrees from the Dept. of Computer Science, University of Erlangen in 1998 and 2003, respectively. Dr. Dressler has been
associate editor-in-chief for IEEE Trans. on Network Science and Engineering, IEEE Trans. on Mobile Computing and Elsevier Computer Communications as well as an editor for journals such as IEEE/ACM Trans. on Networking, Elsevier Ad Hoc Networks, and Elsevier Nano Communication Networks. He has been chairing conferences such as IEEE INFOCOM, ACM MobiSys, ACM MobiHoc, IEEE VNC, IEEE GLOBECOM. He authored the textbooks Self-Organization in Sensor and Actor Networks published by Wiley \& Sons and Vehicular Networking published by Cambridge University Press. He has been an IEEE Distinguished Lecturer as well as an ACM Distinguished Speaker. Dr. Dressler is an IEEE Fellow, an ACM Fellow, and an AAIA Fellow. He is a member of the German National Academy of Science and Engineering (acatech). He has been serving on the IEEE COMSOC Conference Council and the ACM SIGMOBILE Executive Committee. His research objectives include next generation wireless communication systems in combination with distributed machine learning and edge computing for improved resiliency. Application domains include the internet of things, cyber-physical systems, and the internet of bio-nano-things.
\end{IEEEbiography}

\end{document}